# Condensation and the Volatility Trend of the Earth

Katharina Lodders[1], Bruce Fegley[1,*], Klaus Mezger[2], and Denton Ebel[3]

[1]McDonnell Center for Space Sciences and Department of Earth, Environmental, and Planetary Sciences, Washington University, St Louis, MO 63130 USA, bfegley@wustl.edu; [2]Institut für Geologie and Center for Space and Habitability, Universität Bern, Baltzerstrasse 1+3, CH-3012 Bern, Switzerland; [3]Department of Earth and Planetary Sciences, American Museum of Natural History, Central Park West at 79th Street, New York, NY 10024 USA.

## Abstract

This article describes condensation of the elements and use of condensation temperatures to plot and interpret Earth's apparent volatility trend. Major points covered include the following. (1) Updated 50% condensation temperatures ($T_{50}$) for all naturally occurring elements and Pu are tabulated at $10^{-2}$ to $10^{-8}$ bar total pressure for solar composition material. (2) Condensation temperatures are mainly controlled by the Gibbs energy of condensation reactions and also by the Gibbs energy of ideal mixing if elements (compounds) condense in a solution. The additional Gibbs energy change due to non-ideal solution, i.e., activity coefficients ≠1, is a secondary effect. (3) The theoretically correct relationship between condensation temperature and fraction condensed ($\alpha_M$) is derived from mass balance and chemical thermodynamic considerations. The theoretically correct relationship of normalized abundances with condensation temperatures is log $[1/(1-\alpha_M)]$ (for major elements) or log $[\alpha_M/(1-\alpha_M)]$ (for elements in solid solutions) with *inverse* condensation temperatures ($K^{-1}$). (4) The maximum amount of element condensed per $K^{-1}$, i.e., the maximum in $[d\alpha_M/d(1/T)]$ is at the inflection point in the logistic (sigmoid) curve for an element, which is also at (or close to) the 50% condensation temperature. (5) Plots of normalized elemental abundances versus 50% condensation temperatures (volatility trends) are qualitative indicators of elemental fractionations due to volatility because they do not use the theoretically correct and quantitative relationship between condensation temperature and fraction condensed. (6) Volatility trend plots for average elemental abundances in CM, CO, CV, CR, H, L, LL, EH, EL chondrites show different "trends" for moderately and highly volatile elements, which may be linear, curved, a step function, or plateau. A comparison of three abundance sets for CM and CV chondrites shows trends depend on which elements are plotted, which data sources are used, and which temperature range is considered. (7) Proposed mechanisms for volatile element depletion in carbonaceous chondrites and the Earth are reviewed. (8) Some possible implications of volatile element abundances in the bulk silicate Earth are discussed.

## 1 Introduction

Chemical equilibrium calculations of element distribution between gas and solid (liquid) in a solar composition system – aka condensation calculations – have been done for over 70 years and are widely used in cosmochemistry and planetary science (e.g., Urey 1952, Lord 1965, Larimer 1967, 1973; Lewis 1972a,b; Grossman 1972, 1973; Grossman and Larimer 1974, Grossman and Olsen 1974, Wasson and Chou 1974, Boynton 1975, 1978; Palme and Wlotzka 1976, Wasson and Wai 1976, Kelly and Larimer 1977, Wai and Wasson 1977, 1979; Sears 1978, Davis and Grossman 1979, Lewis et al. 1979, Fegley and Lewis 1980, Sears 1980, Fegley 1983, Saxena and Eriksson 1983a,b, Fegley and Palme 1985, Kornacki and Fegley 1986, Palme et al. 1988, Fegley and Prinn 1989, Lodders and Fegley 1993, Yoneda and Grossman 1995, Lauretta and Lodders 1997, Ebel and Grossman 2000, Ireland and Fegley 2000, Lodders 2003, Ebel 2006, Fegley and Schaefer 2010, Wood et al 2019, Lodders and Fegley 2023). Lodders and Fegley (1997) and Ebel (2023) review the history of condensation calculations, and the papers cited above describe many details of the calculations. The major conclusion from this work is that gas – solid (liquid) condensation of the elements (compounds) was the major factor determining elemental abundances in the bodies of the solar system (Larimer 1988). This paper focuses on a subset of solar system bodies, namely chondritic meteorites and the Earth.

The condensation temperatures of elements are the basis for the cosmochemical classification of refractory, moderately volatile, and highly volatile elements (Larimer 1988). Refractory elements condense at higher temperatures than do the major rock-forming elements Mg, Si, and Fe. Moderately volatile elements (MVE) condense at temperatures between the condensation temperatures of the major elements and sulfur (as troilite FeS). Highly volatile elements condense at temperatures below that of sulfur. The exact demarcation temperatures are pressure dependent as shown in Table 1. At $10^{-4}$ bar total pressure, refractory elements condense at T > 1300 – 1360 K, moderately volatile elements condense between 1300 and 660 K, and highly volatile elements condense below 660 K. The cosmochemical classification also uses a fourth group, the siderophile elements, because all the metals are often either depleted (enriched) as a group (Larimer 1988). The different volatility of metals in solar composition gas leads to subdivision into refractory siderophile elements and volatile siderophile elements.

The cosmochemical classification of the elements is often combined with Goldschmidt’s geochemical classification because with a few exceptions the elements have similar geochemical behavior in the early solar system as on Earth (Larimer 1988). Goldschmidt (1923) defined chalcophile, lithophile, and siderophile elements in terms of their affinity for sulfide, silicate, and metallic melts. However, the preference of an element for sulfide, silicate, or metallic host phases during gas – solid (melt) condensation (evaporation) under prevailing temperature, total pressure, oxygen fugacity $fO_2$, and sulfur fugacity $fS_2$ conditions is similar to its preference for these three phases in terrestrial and industrial settings under the same conditions. This point is illustrated by Ellingham diagrams, log $fO_2$ – 1/T, log $fS_2$ – 1/T plots, and predominance area diagrams (isothermal log plots of $fO_2$ – $fS_2$, or $fO_2$ – $P_M$, or $fS_2$ – $P_M$ where $P_M$ is monatomic metal gas pressure) (Ellingham 1944, Kellogg 1966, Ebel et al. 2012). Because the geochemical behavior of an element depends upon temperature, pressure, $fO_2$, and $fS_2$, changes in these variables during nebular condensation, parent body metamorphism, and planetary accretion can change an element’s geochemical behavior.

Some elements, notably Fe, display chalcophile (FeS), lithophile (magnetite, ferromagnesian silicates), and siderophile (Fe metal alloy) tendencies in condensation calculations and in meteorites. Iron is the most abundant metal and is denoted as a siderophile element in the rest of this article. Chromium is another metal with chalcophile ($FeCr_2S_4$), lithophile ($FeCr_2O_4$), and siderophile (Cr in Fe alloy) behavior. It condenses dissolved in Fe metal but is also found in chromite in ordinary chondrites and in Earth's mantle and as daubréelite in enstatite chondrites. The $fO_2$ of solar composition gas intersects the $fO_2$ for the Cr – $Cr_2O_3$ buffer (Holzheid and O'Neill 1995) in the vicinity of the 50% condensation temperature of Cr in Fe alloy (1299 K) at $10^{-4}$ bar pressure. Other elements with mixed geochemical behavior in chondrites and condensation calculations include P and Ga (siderophile and lithophile), and Ag, As, Cu, Pb, Sb, and Sn (siderophile and chalcophile). The phase into which an element condenses and/or occurs in Earth's mantle determines its geochemical classification on the figures in this paper.

Most refractory elements are lithophile elements condensing into oxide and/or silicate minerals. Geochemical analyses of meteorites show that to first approximation, these elements behave as a group with about a 10% - 20% deviation from the mean enrichment (depletion) (Vollstaedt et al. 2020 and this work). As mentioned later there are some exceptions. A few refractory elements are siderophile elements, notably W, Re, Os, Ir, Mo, Ru, Pt, and Rh that condense into refractory metal nuggets (Palme and Wlotzka 1976, Fegley and Palme 1985). The moderately volatile elements (MVE) are a mixture of chalcophile, lithophile, and siderophile elements plus one halogen (F), while the highly volatile elements (HVE) are mainly atmophile (H, C, N, O), chalcophile (Cd, Hg, In, Tl) elements, and halogens (Cl, Br, I).

The abundances of MVE in carbonaceous chondrites and lithophile MVE in the accessible silicate portion of the Earth correlate qualitatively with condensation temperatures; more volatile elements with lower condensation temperatures are more depleted (relative to CI chondrites) than less volatile elements with higher condensation temperatures (e.g., as argued by Wai and Wasson 1977, Palme et al. 1988, Palme and O'Neill 2014, Alexander 2019b). The MVE abundances in non-carbonaceous chondrites such as ordinary and enstatite chondrites are less clearly correlated with condensation temperature. Lithophile and chalcophile MVE can be as abundant as more refractory lithophile and siderophile elements (Alexander 2019a).

The abundances of highly volatile elements (HVE) in chondrites are notoriously difficult to determine and vary with metamorphic grade in non-carbonaceous chondrites (e.g., Lipschutz and Woolum 1988). The elements H, C, N, and O are HVE according to Larimer's (1988) definition and we include them in subsequent plots. The correlation, if any, of highly volatile elemental abundances in meteorites and planetary bodies (such as Earth) with condensation temperatures is less clear. Some analytical data apparently show that subsets of highly volatile elements have constant depletions in CM, CO, CV, and CR carbonaceous chondrites, and in the bulk silicate Earth, relative to CI chondritic abundances, independent of their condensation temperatures (Krahenbühl et al. 1973, Takahashi et al. 1978, Braukmüller et al. 2018, 2019). The temperature dependence of trends for moderately and highly volatile elements (linear, curved, or step function) and the mechanism(s) for any trends remain moot points, are active research areas, and are topics in this review. The focus is on Earth's apparent volatility trend, or more accurately the accessible portion of the Earth commonly known as the bulk silicate Earth (BSE).

The BSE is used in lieu of the bulk Earth composition because the latter is unknown without assumption-dependent modeling of elemental inventories of the Earth's core. However, the BSE composition is also an estimate because most mantle samples originate from the top 200 km of the upper mantle, which is about 27% of the mantle's total mass (section 6.2, Fegley et al. 2020). The majority of the mantle is unsampled, but arguments can be made that the upper mantle composition is representative of the bulk mantle (e.g., see Palme and O'Neill 2014).

This paper is organized as follows. Section 2 presents 50% condensation temperatures for all natural elements at total nebular pressures of $10^{-2}$ to $10^{-8}$ bar. Section 3 mathematically describes the relationship between condensation temperatures and the fraction condensed of an element. Section 4 reviews figures showing volatility trends for CM, CO, CV, CR, H, L, LL, EH, and EL chondrites and the BSE. Section 5 discusses interpretations of the volatility trends given in Section 4 and related topics such as the mechanism(s) for volatile element depletion and isotopic constraints on the origin of Earth's volatile elements. Section 6 summarizes key points and suggests further work to resolve existing controversial issues.

## 2 Condensation temperatures of the elements

Table 1 gives 50% condensation temperatures for all natural elements in solar composition material. The Table 1 values are used throughout this paper. The elements are in atomic number order starting with H (1) and ending at Pu (94). The radioactive elements Tc (43), Pm (61), and Po (84) onward are not included with the exceptions of long-lived Th (90), U (92), and short-lived Pu (94). Plutonium is included because $^{244}$Pu (82 Ma half-life) was present in the early solar system and is important for Pu – I – Xe dating (Hudson et al. 1989, Avice and Marty 2014). The calculations are done as described in Lodders (2003) and Lodders and Fegley (2023). The thermodynamic data sources are given in the Appendix and include the vapor pressure data used for highly volatile elements (H, C, N, noble gases) and a few updates for the alkalis (Na, K, Rb, Cs), halogens (Cl, Br, I), In and Tl. The relative importance of activity coefficients for condensation calculations is discussed with mathematical examples in the Appendix. Table 1 deliberately covers a wide pressure range and interpolation of 50% condensation temperatures ($T_{50}$) between pressures shown can be done with linear fit equations as shown in the Appendix.

A few comments about Table 1 are in order. The 50% condensation temperatures for $H_2$ ice are fictive values at constant assumed total pressure, which drops to 43% of the initial value with half $H_2$ condensed (e.g., see Lewis 1972b). Because of the large amount of hydrogen in solar composition material (with atomic abundance ratios of H/Si $\simeq$ 24,000 and H/O $\simeq$ 2,000) only a tiny fraction of hydrogen is condensed as hydrous minerals and/or water ice at higher temperature. The amount of oxygen in anhydrous rock is likewise insufficient for 50% condensation because the O/(2Si + Mg) atomic abundance ratio is about five. Lodders (2003) discusses the amount of oxygen retained in rock in section 3.4.1 of her paper. Helium never condenses because the condensation temperature of liquid He is below the cosmic microwave background temperature of 3 K. The condensation temperatures for C and N in Table 1 assume complete chemical equilibrium. Possible kinetic inhibition of $CH_4$, $NH_3$, and hydrous silicate formation is described elsewhere (Lewis and Prinn 1980, Fegley and Prinn 1989, Fegley 2000).

Table 1. 50% condensation temperatures (K) at total nebular pressures (bar) of log P =

| Element | −2 | −4 | −6 | −8 |
|---|---|---|---|---|
| H | 10.6 | 7.4 | 5.7 | 4.6 |
| He | <3 | <3 | <3 | <3 |
| Li | 1319 | 1152 | 1003 | 891 |
| Be | 1566 | 1451 | 1349 | 1253 |
| B | 1027 | 945 | 875 | 814 |
| C | 48 | 40 | 34 | 30 |
| N | 139 | 124 | 111 | 101 |
| O | 210 | 182 | 160 | 143 |
| F | 775 | 717 | 699 | 697 |
| Ne | 10.8 | 9.1 | 7.7 | 6.8 |
| Na | 1077 | 978 | 895 | 826 |
| Mg | 1476 | 1330 | 1214 | 1115 |
| Al | 1795 | 1655 | 1533 | 1427 |
| Si | 1451 | 1313 | 1201 | 1111 |
| P | 1421 | 1273 | 1142 | 1029 |
| S | 661 | 661 | 661 | 660 |
| Cl | 460 | 418 | 383 | 354 |
| Ar | 54 | 46 | 41 | 36 |
| K | 1063 | 975 | 899 | 834 |
| Ca | 1683 | 1517 | 1382 | 1269 |
| Sc | 1704 | 1629 | 1505 | 1398 |
| Ti | 1742 | 1581 | 1449 | 1337 |
| V | 1711 | 1429 | 1371 | 1208 |
| Cr | 1490 | 1299 | 1152 | 1036 |
| Mn | 1290 | 1165 | 1049 | 966 |
| Fe | 1531 | 1334 | 1182 | 1063 |
| Co | 1552 | 1352 | 1200 | 1078 |
| Ni | 1552 | 1353 | 1200 | 1078 |
| Cu | 1200 | 1041 | 916 | 817 |
| Zn | 720 | 697 | 668 | 611 |
| Ga | 1206 | 965 | 730 | 665 |
| Ge | 1076 | 887 | 755 | 658 |
| As | 1230 | 1069 | 943 | 839 |
| Se | 693 | 693 | 693 | 693 |
| Br | 466 | 423 | 387 | 357 |
| Kr | 63 | 56 | 50 | 45 |
| Rb | 1161 | 1019 | 913 | 823 |
| Sr | 1682 | 1515 | 1379 | 1265 |
| Y | 1821 | 1607 | 1480 | 1378 |
| Zr | 1902 | 1742 | 1606 | 1489 |
| Nb | 1759 | 1559 | 1390 | 1276 |
| Mo | 1759 | 1592 | 1449 | 1324 |
| Ru | 1712 | 1551 | 1418 | 1306 |
| Rh | 1578 | 1390 | 1256 | 1153 |
| Pd | 1527 | 1324 | 1168 | 1045 |
| Ag | 1158 | 999 | 879 | 785 |
| Cd | 614 | 540 | 481 | 435 |
| In | 560 | 430 | 366 | 315 |
| Sn | 909 | 706 | 619 | 552 |
| Sb | 1148 | 983 | 859 | 764 |
| Te | 811 | 712 | 691 | 641 |
| I | 349 | 316 | 291 | 270 |
| Xe | 81 | 72 | 66 | 60 |
| Cs | 1103 | 857 | 729 | 641 |
| Ba | 1684 | 1490 | 1357 | 1252 |
| La | 1748 | 1583 | 1439 | 1326 |
| Ce | 1725 | 1538 | 1320 | 1266 |
| Pr | 1756 | 1583 | 1437 | 1321 |
| Nd | 1771 | 1613 | 1457 | 1327 |
| Sm | 1748 | 1590 | 1456 | 1328 |
| Eu | 1524 | 1366 | 1218 | 1133 |
| Gd | 1816 | 1619 | 1460 | 1346 |
| Tb | 1819 | 1619 | 1460 | 1346 |
| Dy | 1817 | 1619 | 1460 | 1346 |
| Ho | 1819 | 1619 | 1460 | 1346 |
| Er | 1821 | 1619 | 1460 | 1346 |
| Tm | 1815 | 1619 | 1460 | 1346 |
| Yb | 1699 | 1487 | 1393 | 1296 |
| Lu | 1821 | 1619 | 1460 | 1346 |
| Hf | 1827 | 1683 | 1572 | 1474 |
| Ta | 1774 | 1577 | 1451 | 1327 |
| W | 1988 | 1793 | 1626 | 1480 |
| Re | 2005 | 1821 | 1669 | 1539 |
| Os | 1993 | 1813 | 1662 | 1533 |
| Ir | 1770 | 1603 | 1464 | 1347 |
| Pt | 1585 | 1408 | 1277 | 1173 |
| Au | 1295 | 1065 | 902 | 783 |
| Hg[a] | 273 | 248 | 226 | 208 |
| Tl | 475 | 404 | 361 | 337 |
| Pb | 869 | 730 | 642 | 578 |
| Bi | 879 | 749 | 653 | 580 |
| Th | 1821 | 1619 | 1460 | 1346 |
| U | 1787 | 1610 | 1456 | 1327 |
| Pu[b] | 1740 | 1578 | 1444 | 1331 |

[a]Ideal solution of HgS + HgSe + HgTe in FeS with HgTe being the most important.
[b]From Kornacki and Fegley (1986)

## 3 Relationship of condensation temperatures and fraction condensed

This section describes how the elemental fraction condensed is related to the condensation temperature. This relation is not linear (Larimer 1967, 1973, Kelly and Larimer 1977) and therefore quantitative models using fraction condensed vs condensation temperatures are questionable. The fraction condensed depends on solar elemental abundances, thermodynamic variables, and total pressure. This is illustrated for condensation of siderophile elements into a metal alloy, which is the easiest system to describe, but the general formalism also applies to oxide and sulfide condensation and to more complex gas phase speciation.

The condensation reaction of a monatomic gas into a metal alloy for any element M is

$$\mathrm{M\ (g) = M\ (alloy)} \qquad (1)$$

This reaction applies to refractory siderophile elements such as Re, Os, Ir, Pt, Rh, Pd; to the most abundant siderophile elements Fe, Ni, Co; and to moderately and highly volatile siderophile elements such as Ag, As, Au, Bi, Cr, Cu, and Pb (e.g., see Table 2.3 of Lodders and Fegley 1998). The equilibrium constant for reaction (1) is $K_1$ and the equilibrium constant expression solved for the activity of M in the alloy is

$$a_{\mathrm{M,alloy}} = K_1\, X_{\mathrm{M,gas}} P_{\mathrm{tot}} = X_{\mathrm{M,alloy}} \gamma_{\mathrm{M}} \qquad (2)$$

where $X_{M,gas}$ is the mole fraction of monatomic M gas times $P_T$, the total pressure. The activity, $a_{M,alloy}$ equals the mole fraction of M in the metal alloy, $X_{M\ alloy}$, times the activity coefficient of M, $\gamma_M$, in the alloy. The condensation temperature enters via the equilibrium constant $K_1$.

The fraction condensed, $\alpha_M$, is computed by relating the two mole fractions (one for the gas, one for the alloy) to the bulk composition (here solar). The total number of M atoms is $N_{Mtot}$, which is the sum of atoms in the gas, $N_{Mtot}(1- \alpha_M)$, and $N_{Mtot}(\alpha_M)$, atoms in condensate(s).

On the left-hand side (LHS) of equation (2), the mole fraction of monatomic gas $X_M$ is the number of monatomic gas atoms divided by the sum of *all* other atomic and molecular gases:

$$X_{M,gas} = \frac{\mathrm{M}}{\mathrm{H+H^{+}+H_2+He+CO+CH_4+H_2O+N_2+NH_3+Si+SiO+Fe+\cdots+M}} \approx \frac{M}{\mathrm{H_2+H+He}} \qquad (3)$$

For simplicity, equation (3) denotes numbers of atoms and molecules by the atomic and molecular formulas, e.g., $N_{He}$ = He and $N_{CO}$ = CO. The rest of this discussion uses the same convention, otherwise the equations appear even more complicated. Molecular and monatomic hydrogen and He dominate (>99%) the total pressure in a solar composition system (total pressure = the sum of all partial pressures). At temperatures where condensates become stable the denominator is usually largely $H_2$ and He but at higher temperatures and especially at higher temperatures and lower total pressures, the contributions of monatomic H, $H^+$, and electrons can become dominant and must be included. This is done by considering that the abundances of $H_2$, H, $H^+$, $H^-$, and electrons are coupled though their equilibrium reactions with preservation of mass and charge balance, and can be expressed in terms of each other, e.g., $p(H) = K_H\, p(H_2)^{0.5}$ for the (total pressure-dependent) equilibrium ½ $H_2$ = H. For reference, seven neutral and charged species of hydrogen plus the electron gas are fully considered. Many other ions of other gases are also considered but are not discussed further. Other gases ($H_2O$, CO, $CH_4$, $N_2$, noble gases, SiO, etc.) in the denominator can be neglected to a first approximation here but are fully included in the computations. Mass-balance, charge balance, and total pressure effects are constrained in the calculations for all H-bearing compounds and all compounds of all elements.

The abundance of the major gases $H_2$ and He can be written using their solar atomic elemental abundances as $N_{Htot}$ and $N_{Hetot}$ so that $H_2 \approx 0.5\, N_{Htot}$ and He = $N_{Hetot}$. At low temperatures, the condensation of hydrated silicates, $H_2O$, $CH_4$, and $NH_3$ ices reduces the amount of hydrogen in the gas by a factor of $(1-\alpha_H)$ from the total H abundance, where $\alpha_H$ stands for the fraction of H condensed. Full mass-balance for hydrogen at low temperatures requires solving coupled gas equilibria and solving the distribution of H between gases and condensates. For most practical purposes, the number of $H_2$ molecules is $H_2 \approx 0.5\, N_{Htot}(1-\alpha_H) = 0.5\, N_{Htot}$ for $H_2 \gg [H + H^+ + H^- + \ldots + 2\, H_2O + 3\, NH_3 + 4\, CH_4 + \ldots]$ and $\alpha_H$ is zero.

As noted earlier, the number of M atoms in the gas equals $N_{Mtot}(1-\alpha_M)$. Therefore equation (3) for the mole fraction for monatomic M gas $X_{M,gas}$ can be rewritten as

$$X_{M,gas} = \frac{\mathrm{M}}{\mathrm{H_2+He}} = \frac{\mathrm{N_{Mtot}}(1-\alpha_{\mathrm{M}})}{0.5\ \mathrm{N}_{Htot}+N_{Hetot}} \qquad (3)$$

Using the astronomical epsilon notation for solar abundances for any element relative to H:

$$\varepsilon_i = N_{itot}/N_{Htot} \qquad (4)$$

for He and M in equation (3) gives

$$X_{M\ gas} = \frac{N_{Mtot}(1-\alpha_M)}{\left(\frac{1}{2}N_{Htot}+N_{Htot}\varepsilon_{He}\right)} = \frac{N_{Mtot}(1-\alpha_M)}{N_{Htot}\left(\frac{1}{2}+\varepsilon_{He}\right)} = \varepsilon_M\,\frac{(1-\alpha_M)}{\left(\frac{1}{2}+\varepsilon He\right)} \qquad (5)$$

For reference, solar composition has $\varepsilon_{He}$ of about 1/10. The sum (½+ εHe) can be treated as a constant as long as $H_2$ is the major H-bearing gas, i.e., $H_2 \gg [H + H^+ + H^- + \ldots]$.

The right-hand side (RHS) of equation (2) gives the mole fraction of M in the alloy as the number of condensed M atoms, $\alpha_M N_{Mtot}$ over the sum of all condensed siderophile atoms (= $\alpha_M N_{Mtot}$ plus all other $\Sigma\, \alpha_i N_{i,tot}$) in the alloy:

$$X_{M\ alloy} = \frac{\alpha_M N_{M,tot}}{\alpha_M N_{M,tot} + \sum_{i \neq M} \alpha_i N_{I,tot}} = \frac{\varepsilon_M\ \alpha_M}{\varepsilon_M\ \alpha_M + \sum_{i \neq M} \varepsilon_i \alpha_i} \quad (6)$$

where the RHS follows from expanding with $N_{Htot}$ and making use of equation (4). Now the gas and alloy mole fractions are inserted back into equation (2):

$$a_{M,alloy} = K_1\ \frac{\varepsilon_M}{(0.5 + \varepsilon_{He})} (1 - \alpha_M) P_{tot} = \frac{\varepsilon_M \alpha_M}{\varepsilon_M \alpha_M + \sum_{i \neq M} \varepsilon_i \alpha_I}\ \gamma_M \quad (7)$$

The solar abundance $\varepsilon_M$ cancels out from the numerators on the RHS and LHS of equation (7) but remains in the denominator on the RHS. Rearranging gives:

$$\frac{\alpha_M}{(1-\alpha_M)\left(\varepsilon_M \alpha_M + \sum_{i \neq M} \varepsilon_i \alpha_i\right)} = K_1\ \frac{1}{(0.5 + \varepsilon_{He}) \gamma_M} P_{tot} \quad (8)$$

Before introducing the temperature dependence through the equilibrium constant, it is useful to check the limiting cases for equation (8).

For **pure metal** condensation, the activity coefficient of pure M metal is unity ($\gamma_M = 1$ ) and there are no other elements in the alloy ($\sum_{i \neq M} \varepsilon_i \alpha_i = 0$), and the term containing the fraction condensed for a major host phase element simplifies to

$$\frac{1}{(1-\alpha_M)} = K_1\ \frac{(\varepsilon_M)}{(0.5 + \varepsilon_{He}) \gamma_M} P_{tot} \qquad \text{for M = major element in major phase} \quad (9)$$

This remains a reasonable approximation for the major element during alloy formation as long as $\varepsilon_M\ \alpha_M \approx\ \varepsilon_M\ \alpha_M + \sum_{i \neq M} \varepsilon_i \alpha_I$, which is the case for Fe with much higher solar abundance than most of the other siderophile elements that do not contribute significantly to the total number of condensed atoms in the denominator in equation (8).

The other approximation for equation (8) is for siderophile **trace elements** that are condensing at temperatures where most of the abundant host elements Fe and Ni are already condensed. If the trace element contribution to the total number of atoms in the alloy is negligible such that $\varepsilon_M \alpha_M \ll \sum_{i \neq M} \varepsilon_i \alpha_I$, then $\sum_{i \neq M} \varepsilon_i \alpha_i \approx\ \varepsilon_M \alpha_M + \sum_{i \neq M} \varepsilon_i \alpha_I$ and equation (8) becomes

$$\frac{\alpha_M}{(1-\alpha_M)} \approx K_1\ \frac{\sum_{i \neq M} \varepsilon_i \alpha_i}{(0.5 + \varepsilon_{He}) \gamma_M} P_{tot} \qquad \text{for M = trace element} \quad (10)$$

This simplification does not apply to condensation of highly refractory elements such as Os, Re, and Ir which are similar in abundance, but it works reasonably well for siderophile elements more volatile than Fe and Ni when $\alpha_{Fe} = \alpha_{Ni} = 1$ and $\sum_{i \neq M} \varepsilon_i \alpha_i \approx\ \varepsilon_{Fe} + \varepsilon_{Ni}$.

The **temperature dependence** enters through the equilibrium constant (easily obtained from thermodynamic data compilations) in the form:

$\log K_1 = \Delta S_1/(R \ln 10) - \Delta H_1/(RT \ln 10)$ (11)

where $\Delta H_1$ and $\Delta S_1$ are the enthalpy and entropy of the condensation reaction (1) respectively, for element M, and R is the gas constant. Taking the log of equation (8), inserting for $\log K_1$ in equation (11), and re-arranging gives:

$$\frac{1}{T_{\alpha,M(alloy)}} = \frac{R \ln 10}{\Delta H_1}\left[\frac{\Delta S_1}{R \ln 10} - \log\left(0.5 + \varepsilon_{He,sun}\right) - \log\frac{\alpha_M}{(1-\alpha_M)\left(\varepsilon_M\alpha_M + \sum_{i\neq M}\varepsilon_i\alpha_I\right)} + \log P_{tot} - \log\gamma_M\right]$$
(12)

If needed, the excess enthalpy and entropy of the activity coefficient ($\ln \gamma_M = H^E/RT - S^E/R$) can be combined with the enthalpy and entropy of the reaction.

Setting $\alpha_M = 0.5$ and solving for T gives the 50% condensation temperature of M into the alloy. The 50% condensation temperatures of the different elements into the alloy are inter-dependent through the overall alloy composition given in the term $\sum_{i\neq M}\varepsilon_i\alpha_I$ where each $\alpha_i$ depends on temperature. For each element in the alloy there is an equation like equation (12) or similarly applicable ones that are coupled non-linearly and solved iteratively to obtain the fraction condensed for each element and the composition of the multi-component alloy. With the approximations in Equation (9) the condensation temperatures of pure metals from their monatomic gases are

$$\frac{1}{T_{\alpha,M(pure)}} = \frac{R \ln 10}{\Delta H_1}\left[\frac{\Delta S_1}{R \ln 10} - \log\left(0.5 + \varepsilon_{He}\right) - \log\left(1-\alpha_M\right) + \log(\varepsilon_M) + \log P_{tot}\right] \quad (13)$$

and for a trace element into an alloy:

$$\frac{1}{T_{\alpha,M(trace\ in\ alloy)}} \approx \frac{R \ln 10}{\Delta H_1}\left[\frac{\Delta S_1}{R \ln 10} - \log\left(0.5 + \varepsilon_{He}\right) - \log\frac{\alpha_M}{(1-\alpha_M)} + \log\frac{1}{\left(\sum_{i\neq M}\varepsilon_i\alpha_I\right)} + \log P_{tot} - \log\gamma_M\right]$$
(14)

For all cases the condensation temperature is *not* a linear function of the fraction condensed but instead is inversely proportional to

$$\frac{1}{T_{\alpha_M}} \propto -\log\frac{\alpha_M}{(1-\alpha_M)\left(\varepsilon_M\alpha_M + \sum_{i\neq M}\varepsilon_i\alpha_I\right)} = log\frac{(1-\alpha_M)\left(\varepsilon_M\alpha_M + \sum_{i\neq M}\varepsilon_i\alpha_I\right)}{\alpha_M}$$

per equation (12); also note the minus sign in the inverse proportionality relationship. For the two limiting approximations above for major or trace elements the condensation temperature becomes inversely proportional to

$$log\frac{(1-\alpha_M)}{\alpha_M}$$

for trace elements from equation (14) when using the approximation $\varepsilon_M\alpha_M \ll \sum_{i\neq M}\varepsilon_i\alpha_I$ as done in eqn. (10).

For major elements the condensation temperature is inversely proportional to

$$log(1-\ \alpha_M)$$

as follows from equation (13) with the simplification as done in equation (9).

The general expression for the **fraction condensed with temperature** from equations (8) or (12) is:

$$\log\frac{\alpha_M}{(1-\ \alpha_M)\left(\varepsilon_M\alpha_M+\sum_{i\neq M}\varepsilon_i\alpha_I\right)} = -\frac{\Delta H_1}{T_{\alpha,M(alloy)}R\ln 10}+\frac{\Delta S_1}{R\ln 10}-\ \log\left(0.5+\varepsilon_{He}\right)+\log P_{tot}\ -\log\gamma_M \qquad (15)$$

The conclusion is that the fraction condensed is not a simple function of $1/T_\alpha$.

Figure 1b shows the fraction condensed as a function of $10^4/T_\alpha$ for elements condensing as minerals or dissolving in solid solution in a more abundant mineral. Major element condensation as silicate or another major phase, or trace element condensation into solid solution into silicates can be described by similar equations as illustrated for alloys above. Note the differences between curve shapes and slopes for major elements and trace elements.

The abundant elements (e.g., Fe, Al, Si, Mg, P, S, F, Cl) show steep curves and condense as more or less pure phases e.g., metal alloy, silicates, FeS, $Fe_3P$, F-apatite, and NaCl that also serve as host phases for other minor (trace) elements. In contrast, there are the sigmoid-like shapes for trace elements condensing into solid solutions (e.g., Na, K, B that condense into anorthite at lower temperatures, Ga condensing into Fe-Ni metal and anorthite, Br and I into NaCl). Figure 1a is similar to Figure 6 of Larimer (1967) where fast cooling corresponds to condensation of pure elements and compounds and slow cooling allows formation of solid solutions. Next comes a description of why the curves for the fractions condensed for pure phases containing major elements differ in shape from those of trace elements in solid solutions.

For pure metals (or alloys very rich in element M), using the approximation in equation (9) gives a simplified expression for the fraction condensed from equation (15)

$$-\log\left(1-\ \alpha_M\right) = -\frac{\Delta H_1}{T_{\alpha,M(pure)}R\ln 10}+\frac{\Delta S_1}{R\ln 10}+\ \log\left(\frac{\varepsilon_M}{0.5+\varepsilon_{He}}\right)+\log P_{tot} \qquad \text{for pure metal (16)}$$

Equation (16) can be further approximated for small fractions condensed of a pure phase ($\alpha_M$ < 0.25) where, $-\ln(1-\alpha_M) \approx \alpha_M$ to give

$$\alpha_{M(pure)}/ln10 \approx -\frac{\Delta H_1}{T_{\alpha,M(pure)} R \ln 10} + \frac{\Delta S_1}{R \ln 10} - \log(0.5 + \varepsilon_{He}) + \log(\varepsilon_M) + \log P_{tot} \quad (16b)$$

for small $\alpha_{M(pure)}$.

This is the only approximation that gives a linear relationship of the fraction condensed ($\alpha$) with $-(1/T)$. However, this approximation only applies to a limited amount of a major element condensed and does not apply to trace elements. But it illustrates why the fraction condensed for an element forming a pure phase shows a steep increase with $T_\alpha$ and $1/T_\alpha$ (Figure 1a,b).

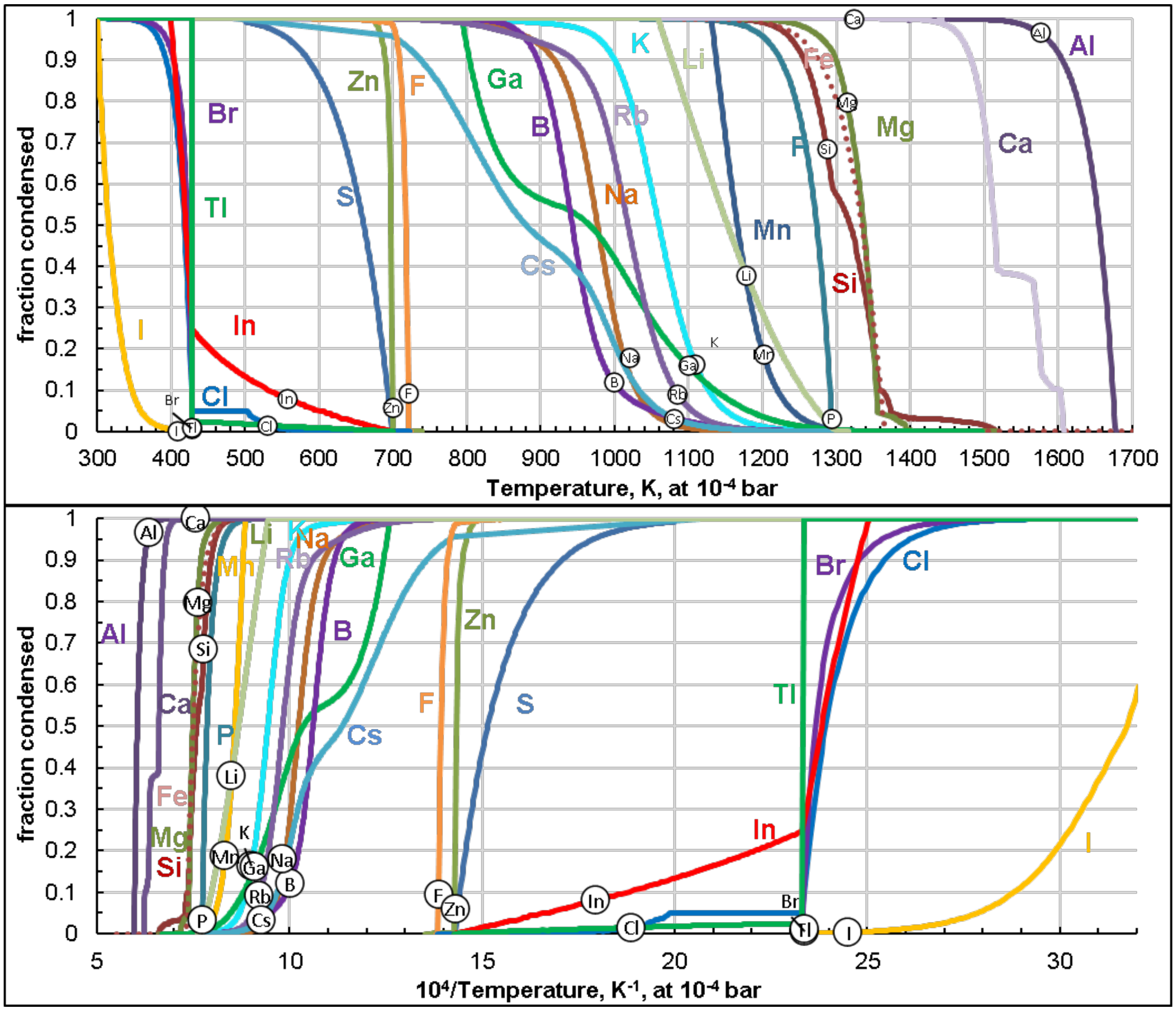

**Figure 1a, b.** Fraction condensed $\alpha_M$ at $10^{-4}$ bar total pressure. (a) on top: $\alpha_M$ plotted versus T. (b) on bottom: theoretically correct representation of logistic function of $\alpha_M$ vs inverse temperature ($10^4$/T scale).

The circles with element symbols in Figure 1 indicate the position of the BSE abundances normalized to CI-chondrites and refractory elements. There are steep increases in the fractions condensed over a very narrow temperature interval for many elements, especially for elements that form major or "pure" phases (e.g., Al, Mg, Si, P, but also F in F-apatite, Cl in NaCl). For these elements, the curves follow log (1-$\alpha_M$). The sigmoid curves are for elements going into solid solution (e.g., Mn, K, B, Br, but also Ga which dissolves in both metal and feldspar) – see equation (22). Indium has a shallow slope until Cl and Br condensation removes halogens from the gas phase. Then the $\alpha_{In}$ slope becomes steeper and all In condenses within a few K. The steep increase in the fraction condensed with decreasing T is the reason why 50% condensation temperatures suggest an apparent trend with temperature. The bottom diagram has the proper logistic function display. The reader should not confuse this treatment with the averaging of $T_{50}$ values done by Sossi et al. (2022) to deduce an average condensation temperature for the Earth.

For trace elements condensing into solid solution, the approximation in equation (10) gives from equation (11),

$$\log\frac{\alpha_M}{(1-\alpha_M)} = -\frac{\Delta H_1}{T_{\alpha,M(alloy)}R\ln 10} + \frac{\Delta S_1}{R\ln 10} + \log\left(\frac{\sum_{i\neq M}\varepsilon_i\alpha_i}{0.5+\varepsilon_{He}}\right) - \log\gamma_M + \log P_{tot} \quad \text{for trace elements (17)}$$

The dependence on the trace element abundance cancels out in the solid solution approximation, and as noted at equation (10), when all Fe and Ni are condensed as metal (no Fe in oxides), $\sum_{i\neq M}\varepsilon_{i,sun}\alpha_i \approx \varepsilon_{Fe} + \varepsilon_{Ni}$.

Thus, for major element condensation, equation (16) gives the dependence with –(1/T) as log (1/(1-α)) but for trace element condensation into a solution, equation (17) shows the dependence is log (α /(1-α)) giving rise to the sigmoid curve shape in Figure 1a,b. The curves for many trace elements in Figure 1b often show inflection points near or at the 50% condensation temperatures. This is the outcome of equation (17), which is a logistic function when re-written for α. From equation (17) in exponential form:

$$\frac{\alpha_M}{(1-\alpha_M)} \approx \frac{P_{tot}}{\gamma_M}\frac{\sum_{i\neq M}\varepsilon_i\alpha_i}{(0.5+\varepsilon_{He})} exp\left(-\frac{\Delta H_1}{T_{\alpha,M}R} + \frac{\Delta S_1}{R}\right) = \frac{1}{u} exp\left(+\frac{\Delta S_1}{R}\right) \exp\left(-\frac{\Delta H_1}{T_{\alpha,M}R}\right) \quad (18)$$

Inverting equation (18) gives:

$$\frac{1-\alpha_M}{\alpha_M} = \frac{1}{\alpha_M} - 1 = \frac{\gamma_M}{P_{tot}}\frac{(0.5+\varepsilon_{He})}{\sum_{i\neq M}\varepsilon_i\alpha_i} exp\left(-\frac{\Delta S_1}{R}\right) exp\left(+\frac{\Delta H_1}{T_{\alpha,M}R}\right) = u\, exp\left(-\frac{\Delta S_1}{R}\right) exp\left(+\frac{\Delta H_1}{T_{\alpha,M}R}\right) \quad (19)$$

The *u* is shorthand for all constants and variables in the preexponential term. Solving for α gives the logistic equation for the cumulate distribution function of the fraction condensed as a function of 1/T:

$$\alpha_M = \frac{1}{1+\frac{\gamma_M}{P_{tot}}\frac{(0.5+\varepsilon_{He})}{\sum_{i\neq M}\varepsilon_i\alpha_i} exp\left(-\frac{\Delta S_1}{R}\right) exp\left(+\frac{\Delta H_1}{T_{\alpha(M)}R}\right)} = \frac{1}{1+u\, exp\left(-\frac{\Delta S_1}{R}\right) exp\left(+\frac{\Delta H_1}{T_{\alpha(M)}R}\right)} \quad (20)$$

For $\alpha_M$ = 0.5 at $T_{50}$ in equations (18) or (19) it follows that

$$exp\left(+\frac{\Delta S_1}{R}\right) exp\left(-\frac{\Delta H_1}{T_{50}R}\right) = \frac{\gamma_M}{P_{tot}} \frac{(0.5+\varepsilon_{He})}{\sum_{i\neq M}\varepsilon_i\alpha_i} = u \qquad (21)$$

(Note the multiple sign changes for the enthalpy and entropy terms from equations (17) to (21)). Substituting equation (21) for *u* into equation (20) gives

$$\alpha_M = \frac{1}{1+exp\left(-\frac{\Delta H_1}{R}\left(\frac{1}{T_{50(M)}} - \frac{1}{T_{\alpha(M)}}\right)\right)} \qquad (22)$$

Equation (22) illustrates that the 50% condensation temperature is not only a convenient volatility measure for a given element but also has practical use for describing the fraction of an element condensed if the 50% condensation temperature is known. The simple logistic function equation (22) from the approximation (!) of trace element condensation, equations (17, 20), has limited use if the enthalpy and entropy of condensation and/or activity coefficient vary significantly with temperature and/or the alloy (or host phase) composition changes significantly with temperature. However, the approximation works fairly well near the 50% condensation temperature.

It is significant that the 50% condensation temperatures are often found around the inflection point of the Hill slope, and for simple cases equation (22) already makes that point. The first derivative of equation (20) for the cumulate logistic function for $\alpha_M$, $d\alpha_M/d(1/T)$ shows that the inflection point corresponds to a local maximum for the amount of an element condensed per inverse temperature step. This important point that the 50% condensation temperatures often correspond to the largest amount of an element condensing per inverse temperature step is shown in Figure 2. Figure 2a shows the negative fraction condensed per degree, and Figure 2b the proper form of differential amount condensed per (1/T). This is another good reason why 50% condensation temperatures are useful proxies for relative volatilities of the elements.

The logistic equation (20) is for trace elements and is easy differentiated to show the significance of the 50% condensation temperature and fraction condensed per (1/T). Proper calculations for every element make use of the general equation (15) which is messy to differentiate. The differential of the re-arranged equation (16) is more practical for estimating the fractional condensation per 1/T interval for elements condensing as pure phases or elements that are major constituents of pure phases. This leads to somewhat different curvatures for $\alpha$, which can be seen for the different elements illustrated in Figure 2b.

The multiple local maxima for Ca in Figure 2 illustrate the importance of multiple stable phases appearing in the condensation sequence for condensation efficiency. Calcium starts to condense as hibonite ($CaAl_{12}O_{19}$), then upon further cooling grossite ($CaAl_4O_7$), melilite etc. each boost the amount of Ca condensing per (1/T). So, each new stable phase is associated with at least one local maximum for differential condensation. A similar situation is shown in Figure 2 for the trace element Ga, which condenses into both metal and feldspar.

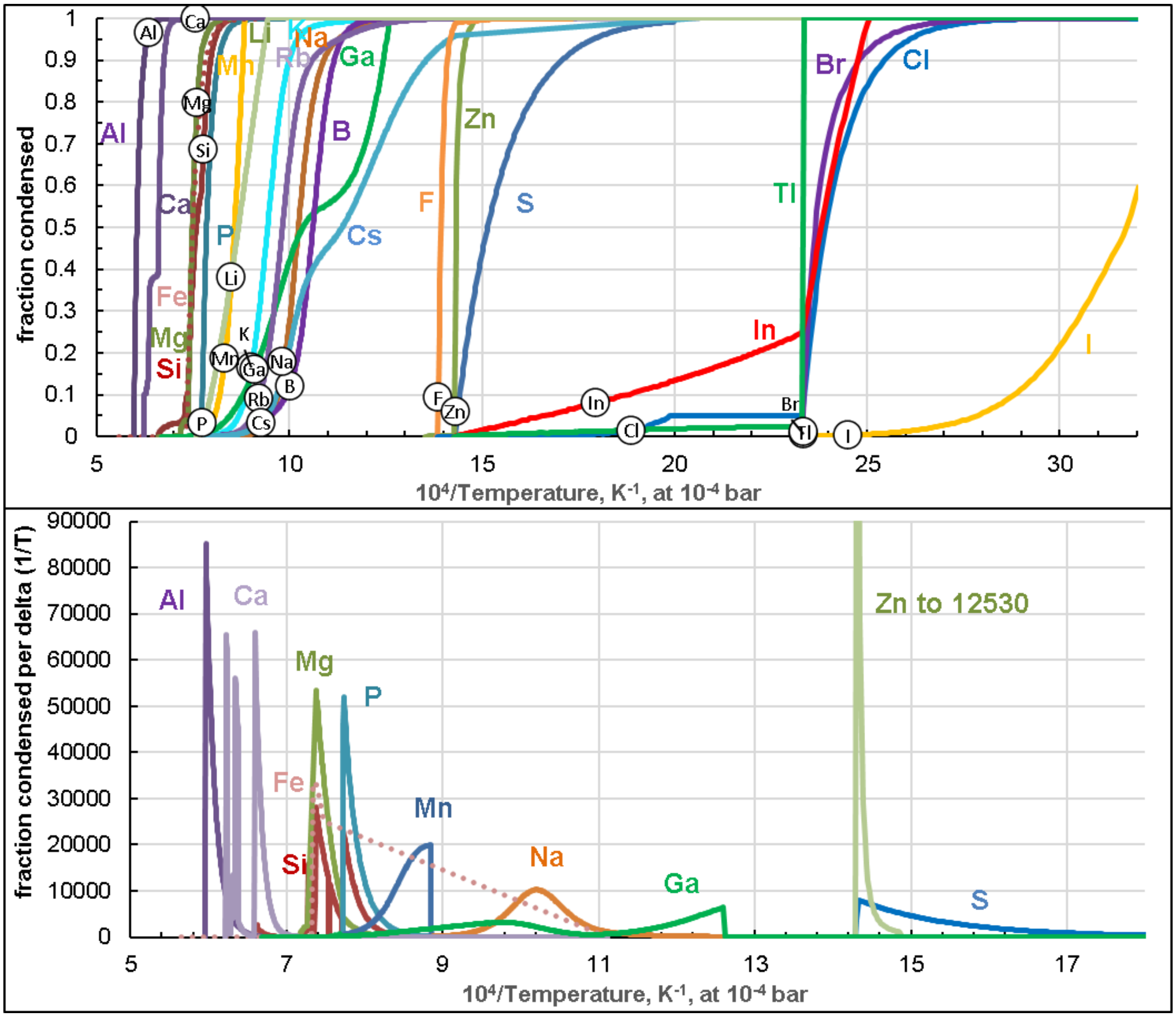

**Figure 2a,b**: (a) on top: The negative of the differential amount condensed per degree. (b) at bottom: The differentials of equations 16 and 20 give the fraction condensed per (1/T) interval.

An important feature in Figure 1b is that the fractions condensed rise steeply along the Hill slope of the sigmoid curves over a narrow temperature (1/T) interval so that smaller and higher fractions condensed occur near the 50% condensation temperature, which means that the temperature (or 1/T) ranges over which most of an element condenses are relatively narrow. For example, at $10^{-4}$ bar, Li has a 50% condensation temperature $T_{50}$(Li) = 1152K and within ±50K of this temperature, 29 to 78% of all Li condenses. Similarly, for Na ($T_{50}$ = 978), and within ±50K of the 50% condensation temperature, 10%-86% of all Na condenses. Fluorine is an extreme case as between the onset of F-apatite stability at 723K and 700K, more than 98% of F is condensed (the $T_{50}$ (F) = 717K). This may explain why the 50% condensation temperature sometimes describes “condensation trends” for fractions condensed other than 50%. But it only works if an element has a steep Hill slope. But to apply this to a set of elements, the elements in question also must condense within a narrow temperature interval with overlapping fraction condensed curves, as there can be only one temperature and pressure for an equilibrium condensation configuration for a given set of elemental abundances (see point (4) in Section 5). This is important for carbonaceous chondrites where observed fractionations often are within one order of magnitude only and could explain some (and only some) observed "trends" with inverse condensation temperatures in these cases.

However, for the BSE, the observed fractionations span about four orders of magnitude (including C and N), so the 50% condensation temperatures are not necessarily close to the temperatures where less than 10% is condensed for elements such as the halogens. The circles in Figure 1 show that the 50% condensation temperatures are far away from the temperatures that correspond to the observed fractions condensed for the CI-normalized BSE abundances. For example, the small amount of iodine in the BSE corresponds to a condensation temperature far from $T_{50}$ for iodine. Similarly, there are two Cl-bearing condensates (chlorapatite and halite). While chlorapatite could explain a small amount of Cl in the BSE, chlorine is not 50% condensed until a much lower temperature when halite (NaCl) forms. Fegley and Schaefer (2010) first showed that the amount of P available for chlorapatite is insufficient to accommodate 50% of all chlorine and that halite condensation is needed to condense all the chlorine remaining in the nebular gas. Similarly, Ga condenses into metal alloy and anorthite, making Ga condensation dependent on the stability of two host phases, and the 50% condensation temperature is far from that for the Ga abundance in the BSE. The indium fraction condensed (Figure 1a,b) increases sharply in the vicinity of the Br and Cl 50% condensation temperatures because their condensation decreases the partial pressures of the two major In-bearing gases, which are InBr (98.8%) and InCl (1.2%). These examples show there is no prescription that fits all elements.

To summarize this section, the popular volatility trend plots of normalized fraction condensed ($\alpha$) versus 50% condensation temperature do not use the theoretically correct relationships discussed above. The fraction condensed for a particular element is not a linear function of temperature as originally shown by Larimer (1967) and also seen in Figure 1. It is only because the 50% condensed point falls in a pseudo-linear region of the sigmoid-shaped condensation curve for an element that plots of normalized abundances versus $T_{50}$ are qualitatively useful.

## 4 Volatility trends for chondrites and the Earth

Considering refractory elements first, Wänke et al. (1974) measured that 29 refractory lithophile and siderophile elements are enriched in a CAI from the Allende CV chondrite by about 20× CI chondritic abundances. Conversely, Wänke et al. found that 13 moderately volatile lithophile and siderophile elements, with lower condensation temperatures, are depleted in the same CAI relative to their CI chondritic abundances. Wasson and Kallemeyn (1988) show refractory lithophile elements are about equally enriched (CM, CV, CO, CR chondrites) or depleted (H, L, LL, EH, EL chondrites) relative to CI, Si normalized abundances. The variable amounts of RLE in chondrites can be explained by the addition (subtraction) of a CAI-like component from CI composition with the Mg/Si ratio balanced by olivine (see Figs. 7.3.5 – 7.3.6 and associated text in Larimer and Wasson 1988). However more recent high-resolution analyses of refractory lithophile elements (RLE) show the RLE are not uniformly present in chondrites and planets (e.g., Dauphas and Pourmand 2015). The Allende CV chondrite (cf. Fig. 6) is the best example because bulk Allende displays a Group II rare earth element pattern (Tanaka and Masuda 1973, Fegley and Kornacki 1984, Stracke et al. 2012, Dauphas and Pourmand 2015).

In Figures 3-13, the RLE show ca. 10-20% deviation from the mean enrichment (depletion) in the volatility trends for carbonaceous, ordinary, and enstatite chondrites. To first approximation, RLE are present in constant proportions, but in detail the different volatilities of the RLE are apparent. Figures 3 to 13 display CI chondrite normalized abundances for all natural chemically reactive elements (color coded as yellow – chalcophile, green – lithophile, blue – siderophile, gray - other) versus temperature to facilitate comparison with published volatility trend plots. The elemental concentration data for CI chondrites are from Lodders et al. (this volume) and from Lodders (2021) and Lodders and Fegley (2023) for the other chondrites.

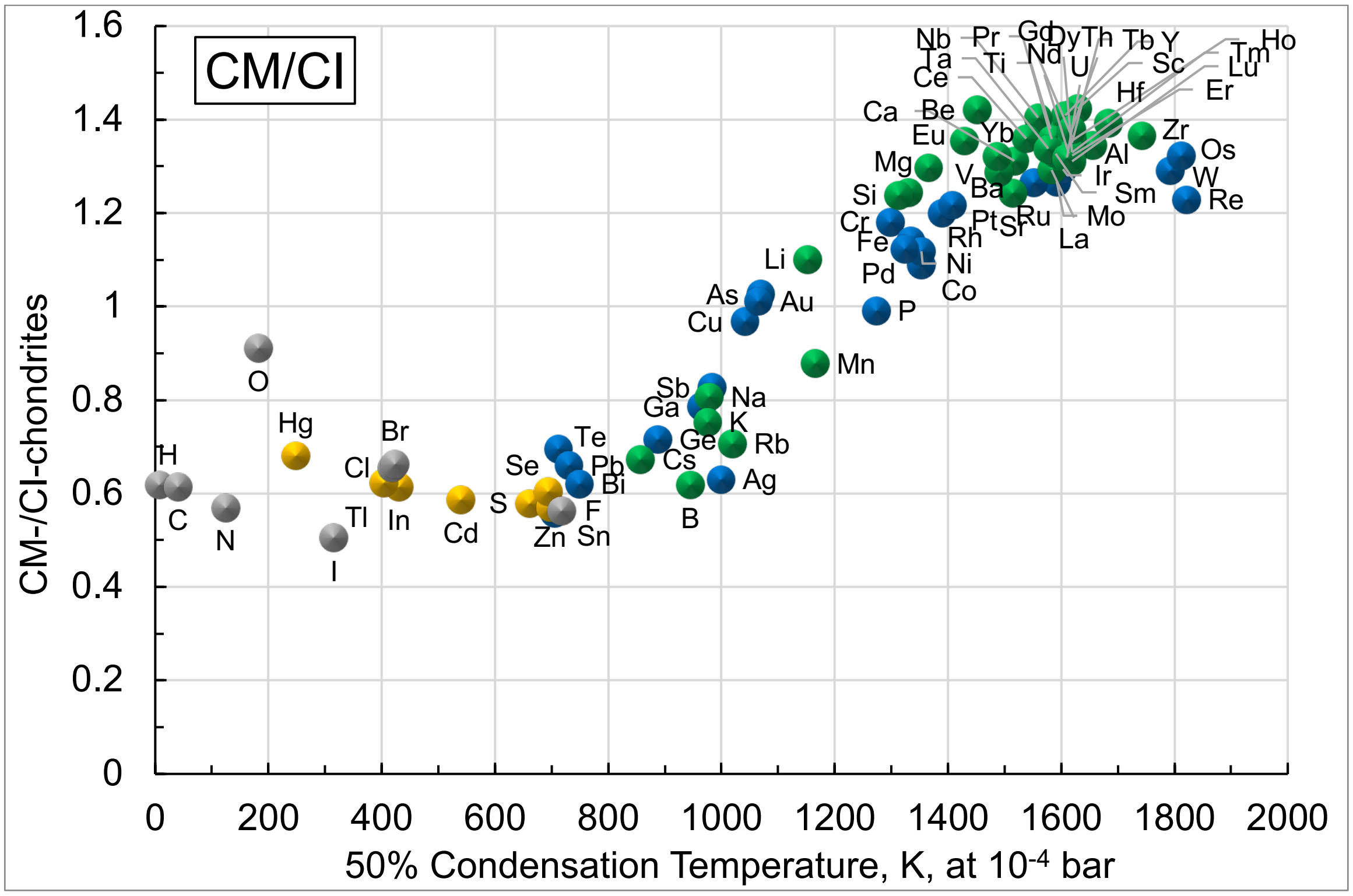

**Figure 3**. CI-chondrite normalized concentrations in CM-chondrites. The CI (418 K) and Br (423 K) points overlap in this and subsequent graphs. There is also some overlap among RLE.

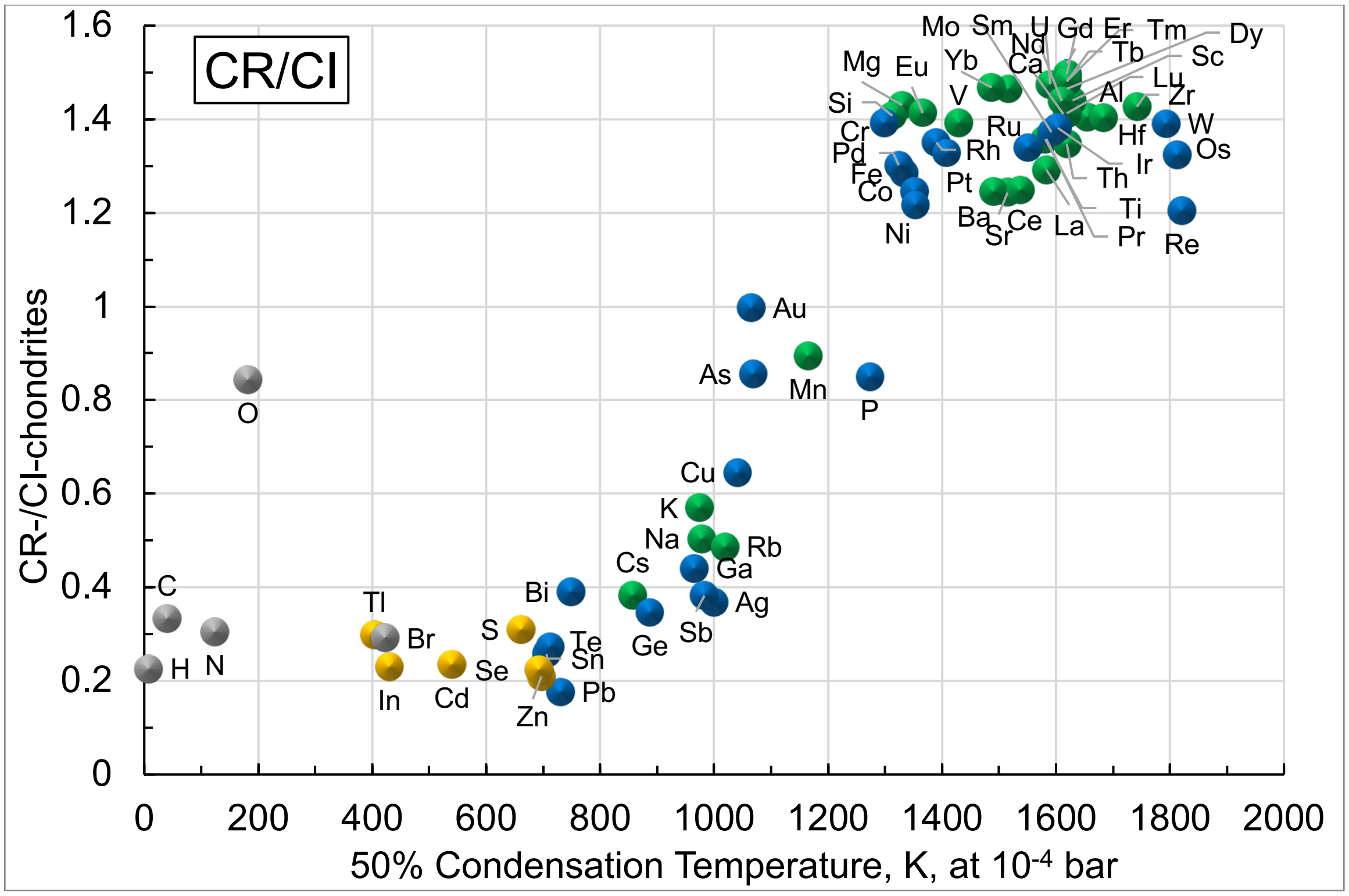

**Figure 4**. CI-chondrite normalized concentrations in CR-chondrites. Note the incomplete coverage for known elemental composition of CR-chondrites.

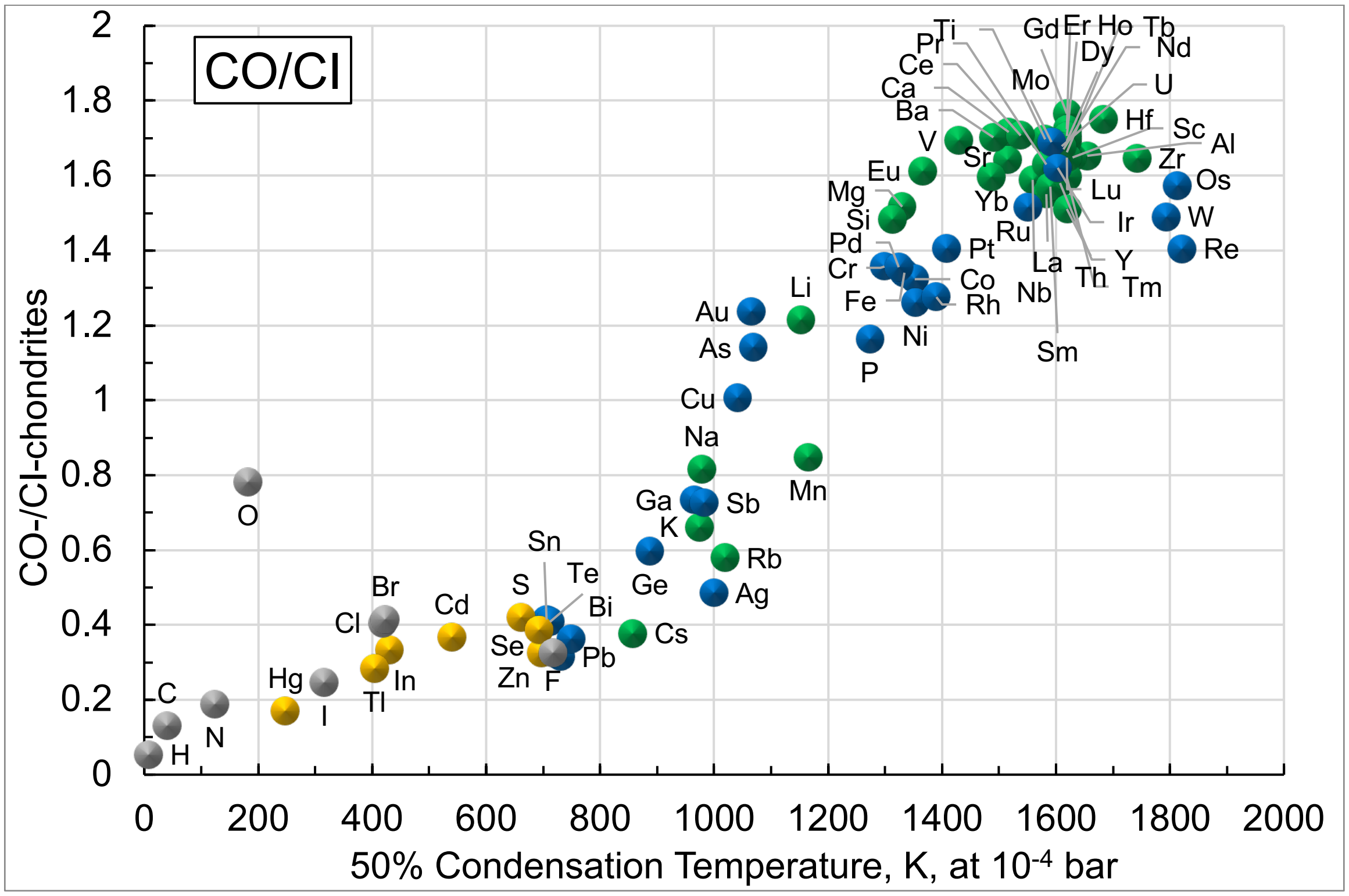

**Figure 5**. CI-chondrite normalized concentrations in CO-chondrites.

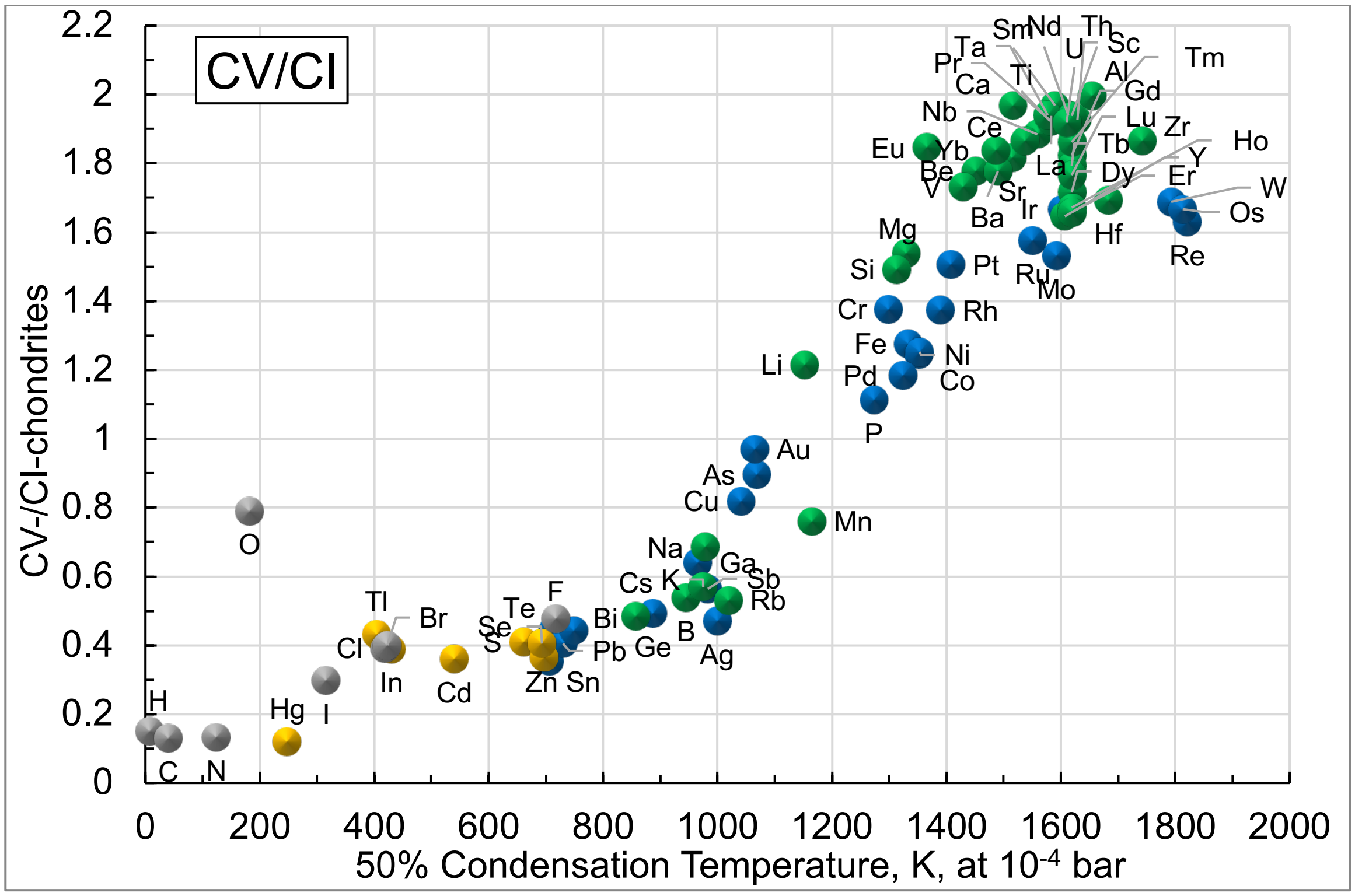

**Figure 6**. CI-chondrite normalized concentrations in CV-chondrites.

Wai and Wasson (1977) plot CI-chondrite normalized abundances of MVE in H and CM chondrites on a logarithmic scale versus 50% condensation temperatures at total pressures of $10^{-4}$ and $10^{-6}$ bar. Wasson and colleagues (Wasson and Chou 1974, Wai and Wasson 1977, Wasson 1977) proposed MVE abundances in CM and H chondrites followed a trend with a continuously negative slope. They suggested this trend was due to incomplete condensation of nebular gas as it cooled and was lost by some process. Wasson's proposed MVE volatility trend for ordinary chondrites was controversial because Anders (1977) interpreted MVE abundances in ordinary chondrites as "a terraced landscape where gently sloping plateaus alternate with steep declines at Cu and Te." Figures 3 to 6 and the volatility trends in Alexander (2019b) show MVE trends for carbonaceous chondrites like that proposed by Wasson and colleagues.

But volatility trend plots for ordinary chondrites in Alexander (2019a) and in Figures 7 to 9 show different behavior for moderately volatile lithophile elements than seen in carbonaceous chondrites. With the exception of B (H, L, LL chondrites) and Cs (H, LL chondrites), moderately volatile lithophile elements plot together with RLE in the upper cluster of points. The steep break (or decline) at about 1000 K is seen best in Figure 7 for H chondrites. Manganese, Li, Na, K, and Rb have higher abundances and Cs, Ga, B, F, and Ga, which condenses in metal and feldspar, have lower abundances. The high alkali abundances in ordinary and enstatite chondrites relative to CI chondrites are well known but not understood (e.g., Goles 1971a,b, Mason 1979). There is likely some connection with the solubility of alkalis in aqueous fluid and the presence of alkali halides in some ordinary chondrites, but further discussion is beyond the scope of this paper (Gast 1962, Lee et al. 2006, Zurfluh et al. 2013, Lodders et al. this volume).

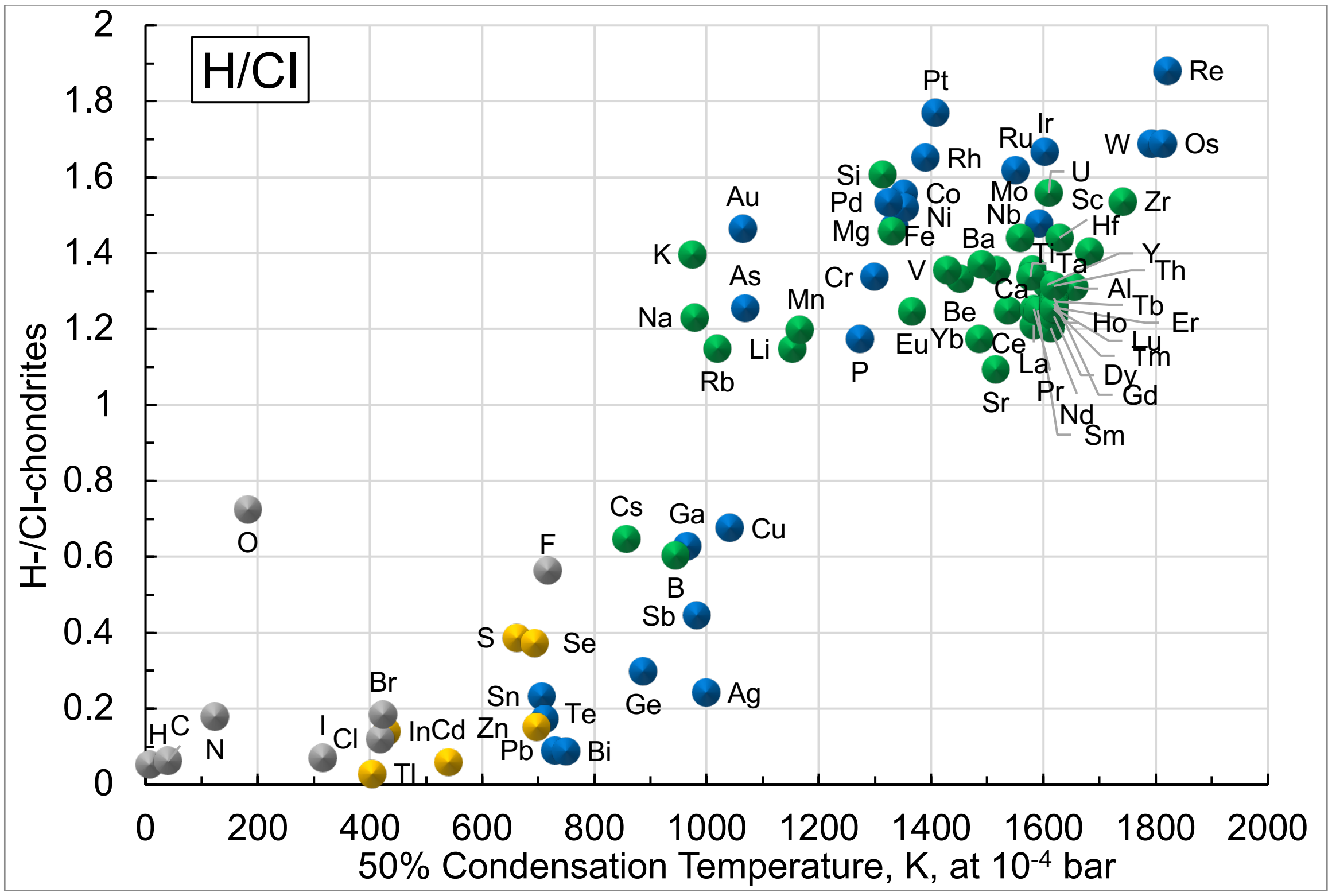

**Figure 7**. CI-chondrite normalized concentrations in H-chondrites.

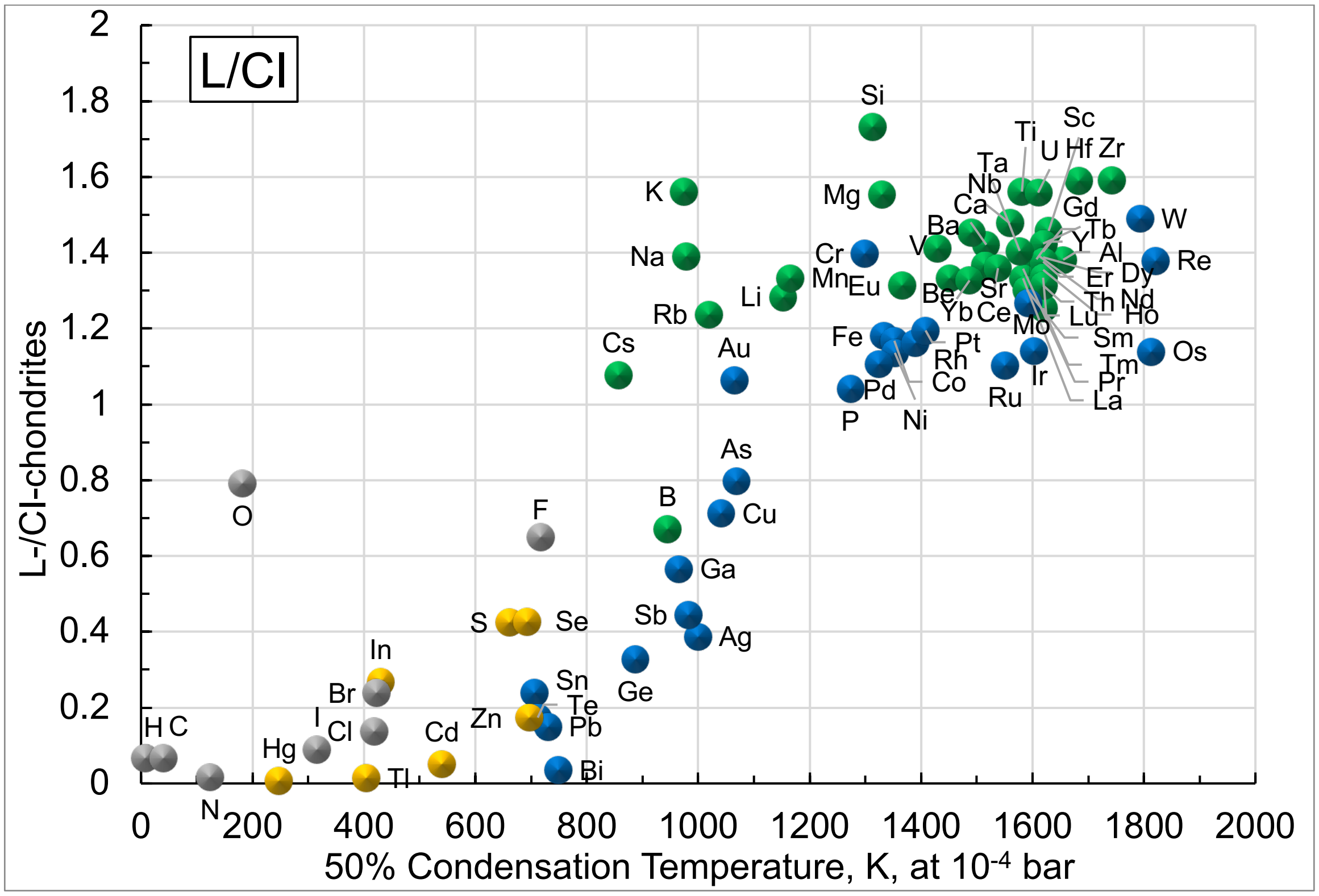

**Figure 8**. CI-chondrite normalized concentrations in L-chondrites.

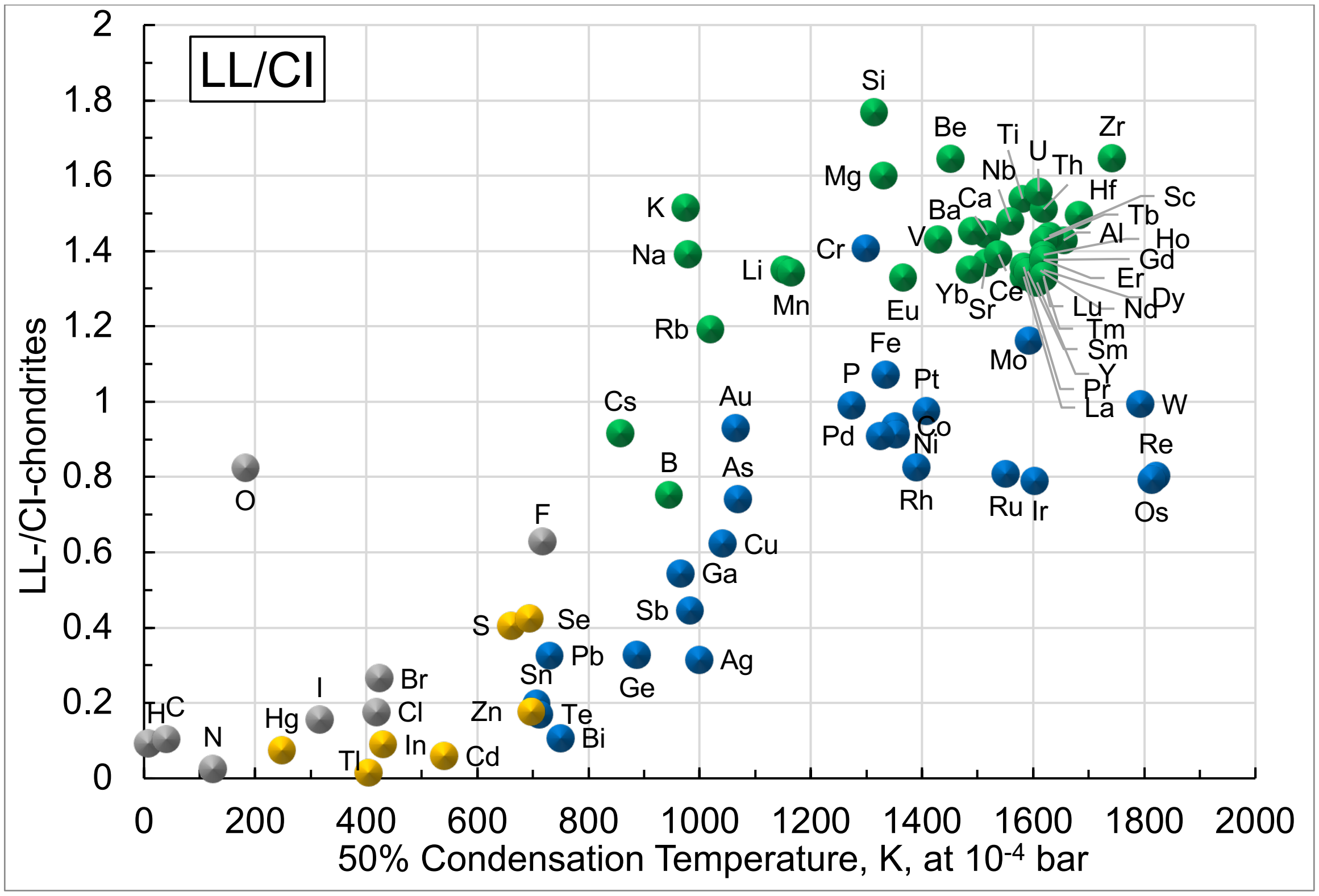

**Figure 9**. CI-chondrite normalized concentrations in LL-chondrites.

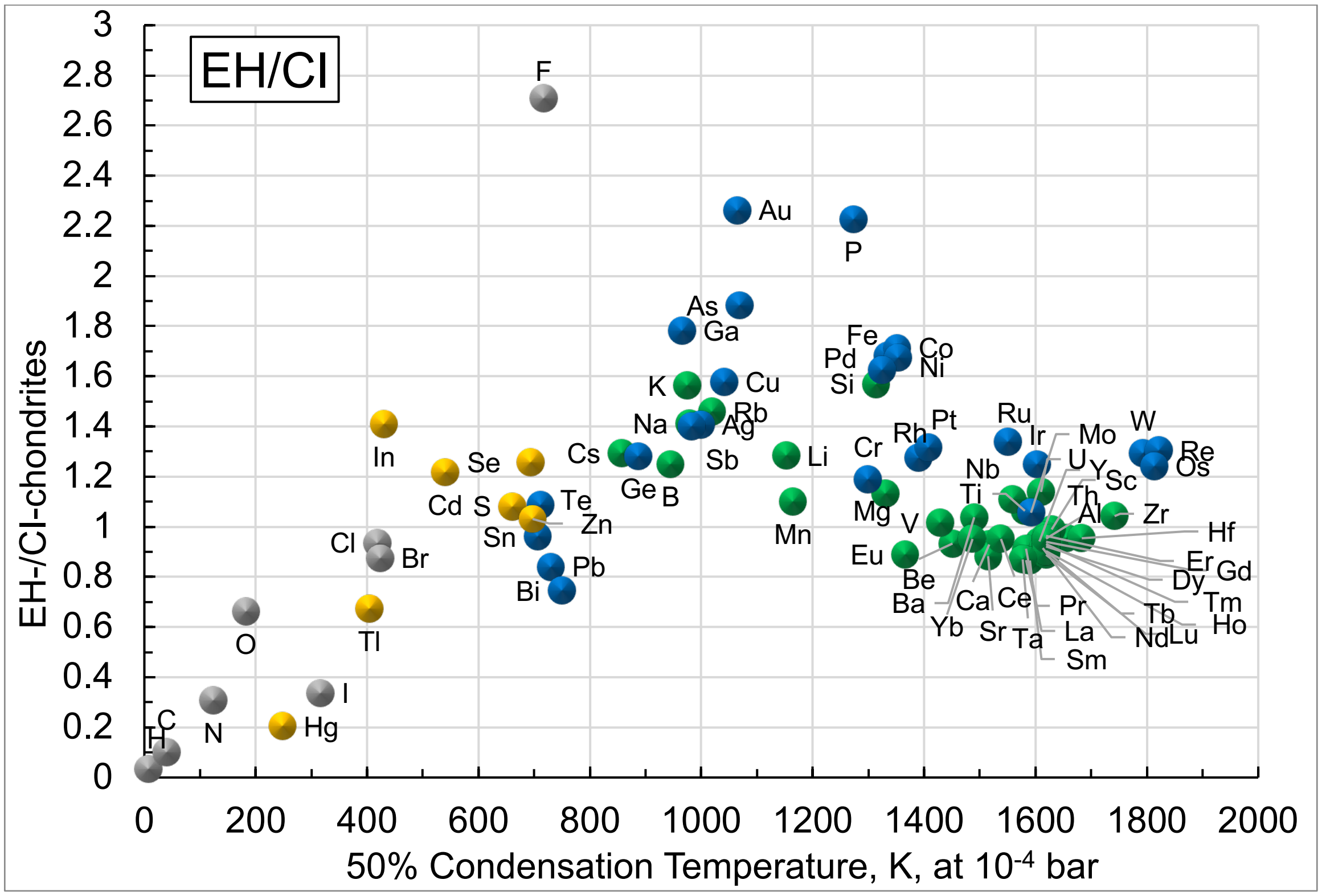

**Figure 10**. CI-chondrite normalized concentrations in EH-chondrites.

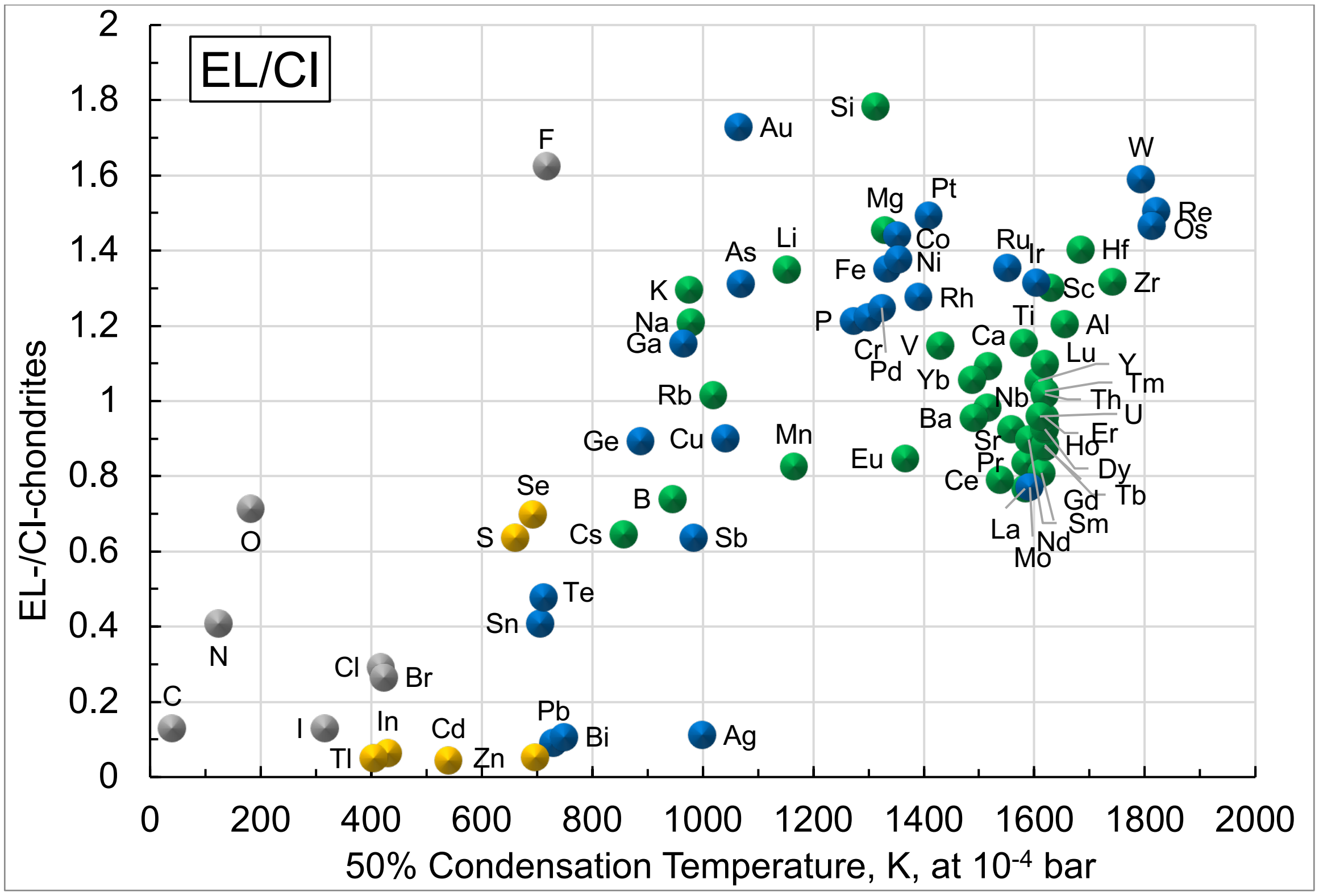

**Figure 11.** CI-chondrite normalized concentrations in EL-chondrites.

The siderophile elements in H chondrites cluster into two groups at higher and lower abundances with a steep break (decline) at about 1000 K. The L and LL chondrites apparently show a volatility trend for siderophile elements more volatile than Au ($T_{50}$ = 1065 K), which is about as abundant as more refractory siderophile elements except the more abundant Cr, Re, and W (L chondrites), and Cr and Mo (LL chondrites).

The EH and EL chondrites in Figures 10 and 11 display complicated trends with moderately volatile lithophile and siderophile elements more abundant than (or equally abundant with) the LRE and refractory siderophile elements. Figure 11 shows the MVE with $T_{50}$ <1000 K (except F, B) are less abundant than the LRE and refractory siderophile elements in EL chondrites. The F abundance in EH and EL chondrites is high. Most of the fluorine analyses were done in the 1960s and on only seven different falls (see Table 2 of Lodders and Fegley 2023). It is important to reexamine the F abundance in enstatite chondrites to check the high F abundance. Alexander (2019a) gives a detailed discussion of volatility trends in ordinary, enstatite, and R chondrites.

Ringwood and Kesson (1977) plotted the increasing depletions of selected chalcophile, lithophile, and siderophile volatile elements in ordinary chondrites, Earth’s mantle, and the Moon relative to their abundances in CI chondrites. They found step patterns for volatile depletions in all cases (their Figs. 1 – 3), but did not use condensation temperatures. Ringwood and Kesson concluded the observed depletions for ordinary chondrites and the BSE correlated at least qualitatively with the condensation temperatures of the elements in the solar nebula. Their depletion sequence for volatile elements in ordinary chondrites disagrees in detail with those plotted with more recent data in Figures 7 to 9 for ordinary chondrites. Ringwood and Kesson (1977) emphasized the lunar depletions were different than those seen for ordinary chondrites and the BSE. Citing several prior papers by Ringwood, they concluded “… that the Moon was ultimately derived from material evaporated from the Earth’s outer mantle and selectively recondensed at some distance from the Earth under unique conditions which permitted further loss of volatile components.”

Palme et al. (1988) presented data showing that the abundances of moderately volatile elements (MVE) in carbonaceous chondrites, differentiated meteorites (e.g., eucrites, SNC meteorites, IIAB and IV irons), and the accessible silicate portions of the Earth and Moon correlate qualitatively with condensation temperatures; less abundant elements are those with lower condensation temperatures. Their review was the basis for subsequent modeling to explain the MVE depletions in terms of condensation during accretion of planetesimals in a cooling solar nebula (Cassen 2001, Ciesla 2008).

The behavior of the highly volatile elements (HVE) that condense at temperatures below that of S (as troilite FeS), was not discussed by Palme et al. (1988). The highly volatile elements include most of the halogens (Cl, Br, I), Cd, Hg, In, Tl, C, N, O, and H. As discussed elsewhere (Lipschutz and Woolum 1988, Lodders 2021, Lodders and Fegley 2023), the abundances of the HVE in chondrites are more uncertain than those of other elements because of possible redistribution and/or loss by metamorphism and/or aqueous alteration for halogen-bearing salts. The temperature dependence, if any, of HVE abundances in chondrites is unclear and still debated actively (e.g., Braukmüller et al. 2018, 2019, Krahenbühl et al. 1973, Takahashi et al. 1978). The HVE in carbonaceous, ordinary, and enstatite chondrites in Figures 3 to 11 show a variety of low temperature “trends”. Oxygen is interesting because its CI normalized abundance gets closer to the HVE trend in the sequence CM, CR ≈ LL, CO ≈ CV ≈ L, H, EL, EH.

Figure 12 compares volatile element abundance trends over the same temperature range using three different elemental abundance data sets for CM and CV chondrites: Alexander (2019b), Braukmüller et al. (2018, 2019), and Lodders (2021) and Lodders and Fegley (2023). The CI abundance data used for normalization are from Lodders et al. (this volume) in all cases. The plots show three different geometries for abundance “trends” that are due to the elements plotted and data used. Different low T cutoffs would further alter the apparent trends.

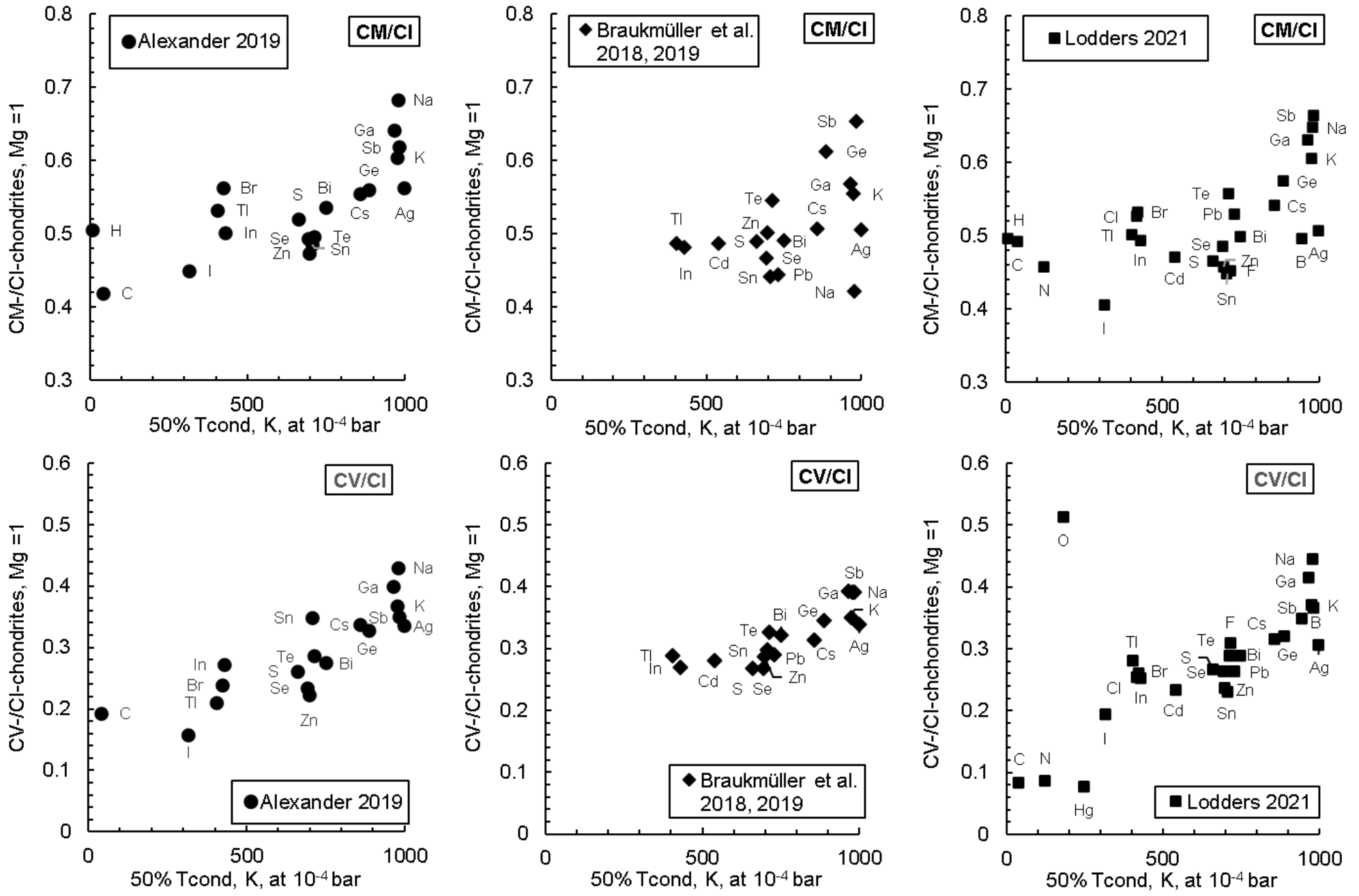

**Figure 12**. Volatile element abundances of CM- and CV-chondrites relative to CI-chondrites and Mg versus condensation temperatures. Data from three references are shown. The values for the abscissa are the same in each graph. See text.

Sun (1982) is apparently the first study to discuss and plot a volatility trend for the bulk silicate Earth using condensation temperatures at $10^{-4}$ atm total pressure. Sun (1982) reached several key conclusions, quoted here: "Above 700 K there are two parallel trends which are defined by lithophile elements (Al, Ca, REE, Ti, Mg, Si, Cr, Mn, Na, K, Rb, F, Zn etc.) and siderophile elements (W, Ni, Co, P, As, Ag, Sb and Ge) respectively. The depletion factor for the siderophile trend relative to the lithophile trend is about 0.085. Within each trend there is a continuous depletion towards lower temperature. A third trend is defined by noble metals (Ir, Os, Re, Pd, Pt and Au) with a depletion factor of about 0.003 relative to CI chondrites. These trends are interpreted in terms of core-mantle differentiation and volatility – controlled processes operating before and during earth accretion." Sun's Fig. 3 is the original volatility trend for the Earth and shows many of the same features as in volatility trends calculated with more recent elemental abundance data for the bulk silicate Earth (e.g., Kargel and Lewis 1993, McDonough and Sun 1995, Palme and O'Neill 2014, Wang et al. 2019). The siderophile element trend he identified involves several of the same elements that apparently show a volatility trend in L chondrites (Co, Ni, P, As, Sb, Ge, Ag). The common siderophile elements in the two trends (L chondrites in Fig. 8 and in Sun's Fig. 3) are both depleted relative to lithophile elements of similar volatility, and for the same reason (but presumably different mechanisms) – metal loss from the sampled material.

With the exception of Wai and Wasson (1977), which focused on ordinary chondrites, all prior work presents volatility trends at $10^{-4}$ bar total pressure. Presumably this was done because no condensation temperatures were available for all elements at other pressures. But whatever the reason, the $10^{-4}$ bar – based volatility trend has become standard in cosmochemistry.

The four plots in Figure 13 show the Earth’s volatility trend computed with 50% condensation temperatures at four different total pressures from $10^{-2}$ to $10^{-8}$ bar to illustrate changes in condensation temperatures, the condensation sequence, and in chalcophile, lithophile, and siderophile element volatility trends.

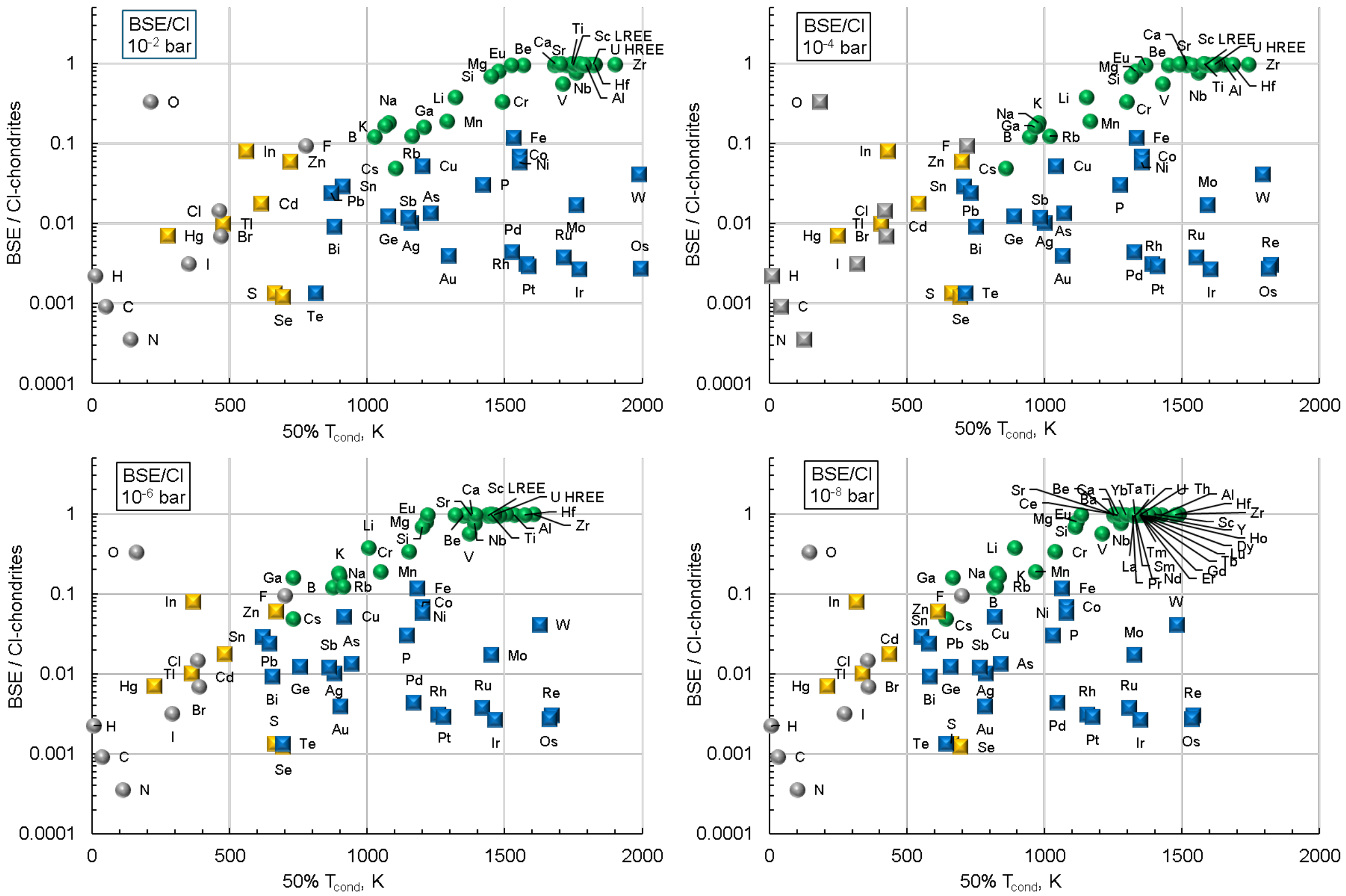

**Figure 13a-d:** BSE abundances normalized to CI-chondrites and refractory elements (Palme and O’Neill 2014, Lodders et al. this volume) versus 50% condensation temperatures from Table 1 at four different assumed total nebular pressures of $10^{-2}$, $10^{-4}$, $10^{-6}$, and $10^{-8}$ bar. All plots are normalized to unity for the refractory lithophile elements. The abscissa is the same in all graphs.

The condensation temperatures are computed for a solar composition gas (Table 1). Thus, normalization of elemental concentrations in the BSE should be relative to the Sun for which atomic abundances are usually given relative to hydrogen. This normalization is impractical, and as generally done, normalization is to CI-chondrites, which are taken as representative for the condensable portion of the elements in the solar composition. The BSE composition is further normalized to the most abundant refractory elements in the BSE (Ca, Al) so the concentrations of refractory elements are at or near unity. This is equivalent to stating that the refractory lithophile elements are fully condensed and present in CI chondritic (i.e., solar) proportions within the bulk silicate Earth, which is the approach taken by Palme and O’Neill (2014). The quantitative relationship of condensation to elemental abundances requires normalizations at unity for the refractories because not more than 100% of an element can condense.

Figures 13a-d illustrate several important points. First, the normalized abundances vary by about four orders of magnitude (considering C and N). This is a much larger range than seen in chondrites. Second, the BSE volatility trend is more similar to that of the carbonaceous than the ordinary or enstatite chondrites (Figures 3 to 11). Third, except for S and Se, all condensation temperatures decrease with decreasing total pressure. The sulfur condensation temperature is constant because its condensation reaction is independent of total pressure [$H_2S$ (gas) + Fe (alloy) = FeS (troilite) + $H_2$ (gas)]. The selenium condensation temperature is constant because Se dissolves as $FeSe_{.96}$ in FeS (Lodders 2003). Fourth, the range of condensation temperatures decreases with decreasing total pressure. At $10^{-2}$ bar, the most refractory element rhenium is 50% condensed at 2005 K and the most volatile element hydrogen is 50% condensed at 10.6 K: a range of almost 1,990 K. At $10^{-8}$ bar, the most refractory element Re is 50% condensed at 1539 K and the most volatile element hydrogen is 50% condensed at 4.6 K: a range of about 1,530 K.

The fifth important point is that the condensation sequence is different as a function of total pressure. The most well-known and important example is the reversal of the condensation temperatures of Fe versus Mg and Si at about $10^{-4}$ bar total pressure. Iron is more refractory than Mg and Si above this pressure and more volatile below $10^{-4}$ bar. The Ga and alkali (Na, K, Rb) condensation temperatures switch with decreasing total pressure with gallium being 130 K more refractory (at $10^{-2}$ bar) to 160 K more volatile (at $10^{-8}$ bar) than sodium. The deviation of In from other volatile chalcophile elements increases with decreasing pressure.

The sixth point is that Figure 13 shows different trends for lithophile and siderophile elements, and noble metals as pointed out by Sun (1982). Sulfur, Se, and Te are grouped in CI-chondritic proportions with abundances lower than the highly siderophile elements Re, Os, Ru, Ir, Rh, Pt, Pd also with CI-chondritic proportions. With decreasing pressure, the sloped trend is better defined with more chalcophile and siderophile elements following the lithophile element trend.

One key question is which elements should be used to define any volatility trend. Should the BSE volatility trend be based only on lithophile elements (Mg, Si, Li, V, Mn, Cr, Ga, B, alkalis, and halogens) or also other elements that appear to follow the same trend such as the atmophile elements and chalcophile elements (Kargel and Lewis 1993, Wang et al. 2019). The chalcophile elements are moderately or highly volatile elements that condense into troilite and/or as another sulfide (i.e., ZnS). Interpretation of the BSE abundances of Zn, Cd, In, Tl, S, Se, Te is complicated because they are affected by volatility, and core and sulfide matte formation. The two heavy Si-group elements Sn and Pb are MVE and depending upon T, $fO_2$, and $fS_2$ behave as chalcophile (PbS, SnS), lithophile ($SnO_2$, PbO) or siderophile elements. How to interpret their BSE abundances? Do Cr and the halogens behave as lithophiles or partly partition into the core (Ringwood 1979, Wänke 1981, Yuan et al. 2023)? Fluorine in the BSE is mainly in the mantle and crust, Cl and Br are mainly in the oceans, and I is mainly in organic marine sediments.

Figure 13, which is widely used, is plotted with the incorrect functional relationship between fraction condensed and 50% condensation temperature. Inverse temperature, not temperature, is the correct independent variable to use. This was done by Humayun and Cassen (2006). The reason is simple – this is the temperature dependence of the equilibrium constants used to compute condensation temperatures. Larimer (1967, 1973) was probably the first person to realize this. When inverse temperature is used, the volatility trend looks completely different as shown in Figure 14 below. The apparent linear diagonal trend is now a convex (or parabolic?) curve with decreasing temperature. Arguments based upon the shape of the volatility trends plotted as in Figure 13 may not be meaningful.

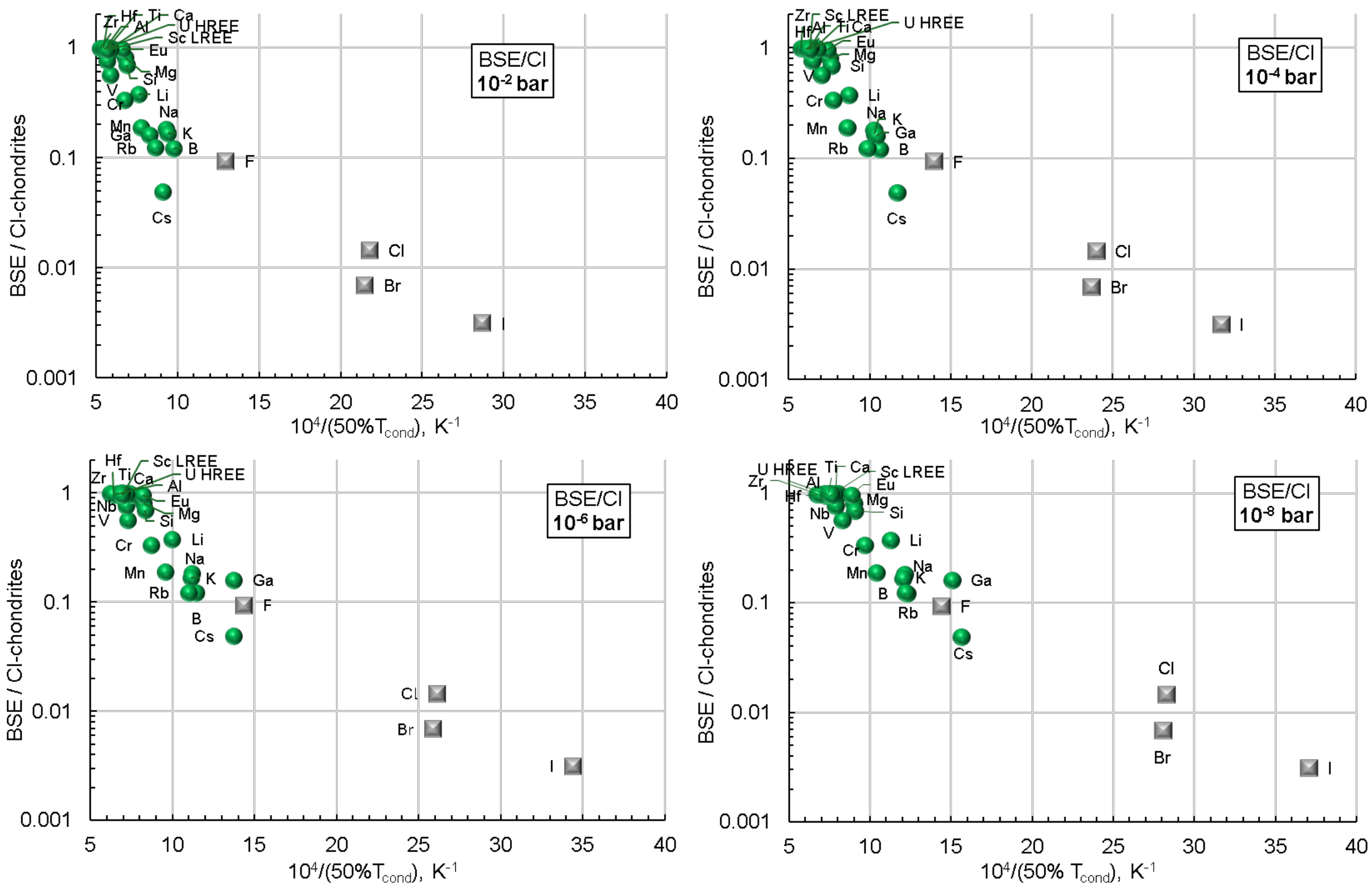

**Figure 14a-d**: The same data as in Figure 13 but plotted versus inverse 50% condensation temperatures. Chalcophile and siderophile elements are not shown.

But even Figure 14 is formally incorrect because the fractions condensed ($\alpha$ values) for major elements and trace elements dissolved in solid solution have different functional dependence upon temperature. The reasons for the apparent trend are related to the steep Hill slope of the logistic function of the fraction condensed with inverse temperature, as described in Section 3.

# 5 Discussion

**The premise of deriving a volatility trend and the nature of the problem.** Figures 3 to 13 show volatility trends for chondrites and the bulk silicate Earth (BSE) that are plotted as CI-chondritic normalized elemental abundances versus the 50% condensation temperatures of the elements from a solar composition gas, usually at $10^{-4}$ bar total pressure. Often the observed elemental abundances are doubly normalized to CI chondrites and a reference element (e.g., Si, Mg, Ca, Al, Yb), e.g., as done in Figure 6-1 of Ebel et al. (2018). Unfortunately, in many studies that attempt to illustrate or to quantify volatility trends the observed abundances are not related to the correct functional dependence on condensation temperatures. Four underlying assumptions often made are as follows.

(1) The source material for condensation was *solar* in elemental composition. Using CI chondrites for normalization instead of solar photospheric composition is done for practical reasons because not all elements can be quantitatively measured in the sun, and the CI chondrites are the best available proxy for the solar condensable elements where comparison can be made (see e.g., Lodders (2020) and Lodders et al., this volume). The CI-normalized abundances are a measure of the elemental fractions condensed from an originally solar-composition-like gas. Normalizing to CI chondrites does not imply that the Earth or meteorite parent bodies formed from material of the same mineralogy or oxidation state as CI-chondrites.

(2) The linear or logarithmic elemental fractions condensed, $\alpha_M$, are directly proportional to the 50% condensation temperatures. This assumption is incorrect. As explained above and in less detail by Larimer (1973), the correct proportionality of normalized abundances is log $(1-\alpha_M)$ for major elements or log $[(1-\alpha_M)/\alpha_M]$ for trace elements in solid solutions with *inverse* condensation temperatures. For several elements, a more complicated dependence of log $f(\alpha_M)$ with their inverse condensation temperatures applies.

(3) Although widely assumed, a total pressure of $10^{-4}$ bar is model specific. Temperature and pressure varied with radial distance from the proto-Sun, height above the nebular midplane, and with temporal evolution of the nebular accretion disk (e.g., Cameron and Fegley 1982, Ruden and Lin 1986, Cameron 1995, Cassen 2001, Ciesla 2008). Lewis (1974) derived a total nebular midplane pressure of $10^{-4}$ bar at 1 AU on the basis of planetary compositional data, an equilibrium condensation model, and particular adiabatic thermal profiles in the solar nebula. In general, lower total pressures are associated with lower temperatures in the nebular midplane.

As a consequence, more volatile elements will condense at lower total pressures than more refractory elements. The net effect is a slightly lower condensation temperature than on an isobaric profile and a more shallow slope in conventional volatility plots like Figures 3-13. For example, at $10^{-4}$ bar, the moderately and highly volatile elements condense from 1324 K (Pd $T_{50}$) to 248 K (Hg $T_{50}$). The change in pressure with temperature for an ideal gas adiabat with solar abundances of $H_2$ and He in this temperature region is given roughly by

$$\frac{dlogP}{dlogT} = \frac{(C_P/C_V)}{[(C_P/C_V)-1]} = \frac{\gamma}{(\gamma-1)} \sim 3.4$$

The change in pressure is about two orders of magnitude pressure drop from 1324 to 248 K. Interpolation of the condensation temperatures in Table 1 can be done to estimate condensation temperatures for thermal profile models of the solar nebula (see the Appendix).

(4) It is often assumed that a physical basis exists for application of 50% condensation temperatures to define volatility trends. These element-specific temperatures describe the temperature, $T_{50}$(M), where 50% of a given element M is condensed and the other half remains in the gas at constant total pressure. Thus, a relative volatility temperature scale can be derived for a given thermal profile such as an isobar or an adiabat. However, there is no inherent physical-chemical relationship between the observed normalized elemental abundances for any particular meteorite parent body (or the bulk silicate Earth) and the relative volatility of the elements. Inference of any such relationship must rely on imposition of a causal model. There is no fundamental reason to expect a relationship between 50% condensation temperatures at any single total pressure and elemental abundances in the BSE or any meteorite parent body.

Examining point (4) in more detail, if equilibrium condensation from a solar composition gas were the *sole* mechanism for establishing elemental abundances in chondrites or in the Earth, in principle there should only be one temperature and one total pressure for which the observed fractions condensed can be reproduced by full equilibrium condensation from a solar composition gas. This approach is analogous to finding a correspondence temperature that matches all gas equilibria in a volcanic gas (e.g., Symonds et al. 1994). But a basic argument indicates that the Earth's volatility trend cannot be matched at any single P, T point.

Consider the temperature where a refractory element is 50% condensed. Then at the same temperature, the fraction condensed of a volatile element with a lower condensation temperature obviously has to be less (or even be zero). Vice versa, at the temperature where the volatile element is 50% condensed, the refractory element would have a fraction condensed larger than 50%. Urey (1954) first identified this problem, which Larimer and Anders (1967) described succinctly: "Vapor pressure is a steep function of temperature; hence a partial depletion can be achieved only over a narrow temperature range, differing from element to element. Outside this range, all-or-nothing fractionations will be the rule, with complete retention or complete loss." Figure 4 of Palme and Wlotzka (1976) shows an example of vapor pressure control; the more volatile siderophile elements such as Au are simply not present in refractory metal nugget grains condensed at high temperatures.

**Accretion of the Earth.** The chemical and isotope composition of the bulk Earth is distinctly different from all known meteorite compositions. The major differences are observed in the abundances of the moderately volatile elements relative to the refractory elements (e.g. Palme and O`Neill, 2014). CI-chondrite normalized elemental abundances in the bulk silicate Earth (BSE) displayed versus condensation temperatures of the respective elements (Figures 13 and 14), define a *rough* depletion trend for the volatile elements, e.g., Ebel et al. (2018) computed a $r^2$ coefficient of 0.33 for the BSE volatility trend. The apparent trend observed for the BSE composition is the result of several chemical processes starting from differentiation during the formation of solids in the solar nebula to planetesimal and planetary accretion to planet-wide differentiation. This path of chemical differentiation from solar nebula to a differentiated planet can be reconstructed from the elemental abundances in the BSE. Traditionally this volatile element depletion trend in the BSE is interpreted as indicating different extents of element condensation during cooling from the hot solar nebula (e.g. McDonough and Sun, 1995). But these models do not consider secondary processes that modify elemental abundances in a planetary body after primary accretion, particularly core formation.

In order to explain the composition of Earth in comparison to different meteorites, qualitative and semi-quantitative accretion models have been presented (e.g., Ringwood 1979, Barshay 1981, Wänke 1981, Dreibus and Wänke 1984, Wänke et al. 1984, Wänke and Dreibus 1988, Lewis 1988, Lodders 1991, 2000, Lodders and Fegley, 1997; Albarède, 2009; Schönbächler et al., 2010; Warren, 2011; Liebske and Kahn, 2019, Sossi et al 2022). These models use element abundances and/or isotope compositions to derive accretion mechanisms for the Earth and suggest chemical and isotope compositions for these components (e.g. Dauphas et al., 2017). One key result is that that the chemistry for the Earth is unlike the composition of any single meteorite group or a simple mixture of different known groups. The Earth has an end member composition for certain element abundances and isotope compositions (e.g., Dauphas et al., 2014; Fischer-Gödde et al., 2013; 2015, Mezger et al., 2020). The accretion models can be separated in two distinct groups. One type of model assumes continuous accretion of planetesimals to achieve Earth's current size (e.g. Sossi et al., 2022). The other model interprets the Earth as the result of planetesimal accretion until ca. 90% of its current size was achieved, and then the last 10% were added, possibly from a different feeding zone, as a giant impactor (e.g. Albarède, 2009; Schönbächler et al., 2010; Mezger et al. 2021). Only the latter model accounts for the origin and peculiar composition of the Moon and is the preferred model.

The abundances of different chemical elements in BSE can be estimated with good precision for most elements (McDonough and Sun, 1995, Palme and O`Neill, 2014, Liebske and Khan, 2019). There are somewhat larger uncertainties for the highly incompatible elements (e.g. Mezger et al 2020) and the highly volatile elements, particularly halogens (e.g., Lodders and Fegley 2023). As shown in Figure 15, the CI-chondrite normalized element abundances in the BSE define some moderately well-constrained trend when plotted versus condensation temperature. The moderately volatile elements scatter around a slope declining with condensation temperature. A common interpretation of this trend is that it represents different degrees of volatile element depletion; either due to incomplete condensation from the solar nebula (e.g., McDonough and Sun, 1995) or later volatile loss during accretion (e.g., Halliday, 2004). A consequence of incomplete condensation or partial evaporation is fractionation of stable metal isotopes. However, no such isotope variations are observed for moderately to strongly volatile elements that are depleted in BSE relative to the chondritic reservoir. Thus, based on the lack of stable isotope variations, the volatile element abundances in BSE are not the result of incomplete condensation nor partial evaporation (Mezger et al. 2020, 2021).

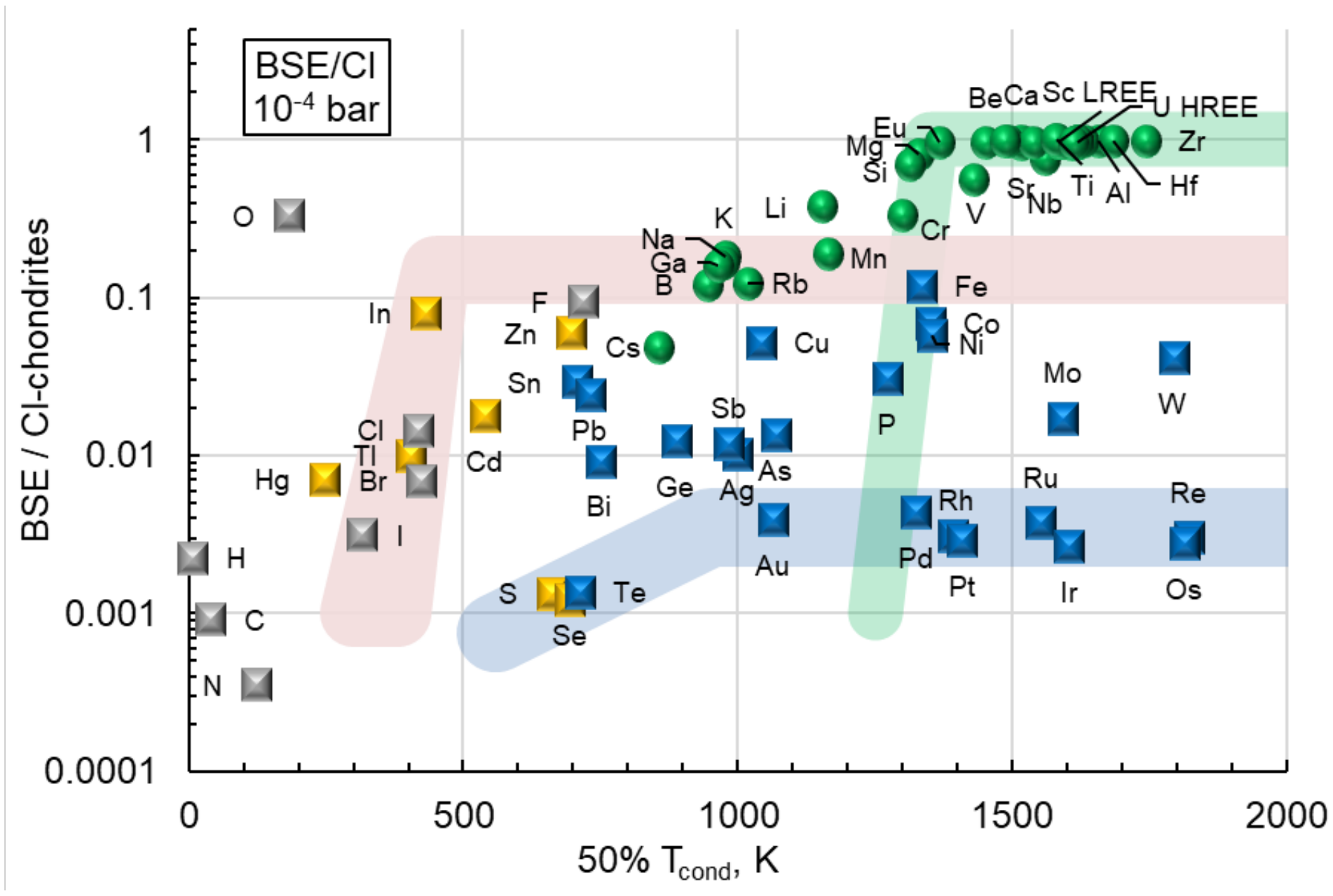

Figure 15. Deconvolution of the apparent BSE volatility trend, modified after Mezger et al. (2021). The green curve is the volatility trend of the proto-Earth, about 85% of the BSE. The red curve is the volatility trend of the Moon-forming impactor Theia, about 15% of the BSE. The blue curve is the volatility trend of the late veneer, about 0.4% of the BSE.

The slope of the apparent element depletion trend in BSE shown by the moderately volatile elements is gentler than that derived from a cooling system where at a point in time the solids are removed from the gas phase, as expected during the initial stages of dust accretion and subsequent planetesimal formation. This relationship is demonstrated by the calculations in section 3. In this case elements that condense above the temperature of dust removal will show chondritic abundances and elements that condense below this threshold temperate would not have condensed significantly. Only elements condensing within a narrow temperature range around the temperature interval where gas and solid fractionate, will condense to different degrees. As demonstrated by Figures 1 and 2 the element pattern in rocky materials will show a steep depletion trend for volatile elements with the depletion starting around the temperature of gas-solid separation. Elements above this temperature interval are essentially completely condensed, while elements that condense below this interval are depleted (absent) in the solids that form planetesimals. The condensation accretion model of Barshay (1981) found this behavior for rocky elements and volatiles (S, $H_2O$, and C).

The element abundances in the BSE show this type of steep depletion pattern, if only the lithophile elements, which are unaffected by extraction of a metal or sulfide melt from the silicate reservoir, are considered. Many moderately volatile elements are siderophile and/or chalcophile and their abundances in the BSE are not *primary* signatures of accretion. If only the lithophile elements in the BSE are considered the element pattern shows two distinct groups with lithophile element abundances close to CI chondrite relative abundances (Figure 15). The first group includes the refractory lithophile elements (red line bracket) and the second group is defined by the moderately volatile lithophile elements (blue line bracket). The abundances of the MVE are ca. 15% that of the refractory lithophile elements. Thus, lithophile elements in the BSE define a "step" pattern with two distinct flat regions rather than continuously declining element abundances with decreasing condensation temperatures of the relevant elements. This relationship indicates that the BSE consists of two components. The first component, i.e. proto-Earth, had chondritic abundances of refractory elements and was strong depleted in moderately to highly volatile elements. The element pattern in Figure 15 shows a depletion around the temperature interval where Mg, Si and Fe condensed. The second component that built the Earth was the Moon-forming impactor Theia. This smaller body (ca. 10% the mass of the current Earth, or 15% of the BSE mass) had chondritic abundances of the refractory elements and the moderately volatile elements but was depleted in highly volatile elements (see Figure 15) (Maltese and Mezger, 2020, Mezger et al., 2021). The combination of these two major planetary bodies and subsequent core formation accounts for the element patterns recorded in the present day bulk silicate Earth. The abundances of the highly siderophile refractory elements as well as S, Se and Te are the result of a late veneer after core-mantle differentiation and account for ca 0.4% of the Earth.

A key feature of the BSE (Fig. 15) is the "step" pattern for elements, with high and chondritic abundances of refractory lithophile elements and low, but also chondritic relative abundances of moderately volatile lithophile elements. The chalcophile and siderophile elements fit these patterns, but their abundances were modified by core and sulfide matte formation. The "step" pattern results from accretion of material extremely depleted in volatile elements, with the end of efficient condensation during cooling of the solar nebula having occurred at different temperatures for different planetary bodies. The precursor material for the proto-Earth included refractory elements with 50% condensation temperatures $\geq T_{50}$ for Mg, Si, and Fe. The precursor material of Theia contained material with $T_{50}$ values down to those of the halogens (Fig. 15). For Vesta such a steep pattern is also recorded with the transition region being around Mn (Fig. 1 in Sossi et al. 2022). Such element patterns are exactly as predicted for incomplete condensation. Due to the fact that the condensation interval for 95% condensation of an element is generally less than 50 K (Fig. 2), and distinct from other elements, flat chondrite normalized element patterns for refractory elements and near absence of volatile elements with a narrow temperature interval with moderate element depletion are expected (see Fig. 15 for the element patterns for proto-Earth and Theia). Mixing of different components with such element patterns gives an overall element depletion pattern with strongly different abundances, particularly of the more volatile elements. Such a mixture also accounts for the lack of stable isotope fractionation of volatile (i.e. depleted) elements in the BSE.

As a consequence, the element abundances in the BSE plotted against their respective condensation temperatures are not the sole result of processes in the solar nebula (i.e. incomplete condensation of the elements) nor planetary accretion processes (i.e. evaporation during accretion) but are a result of mixing of two planetary bodies with distinct element abundances that was followed by core formation. The gentle trend is the result of primary element abundances and the fact that many moderately volatile elements, that define the gentle depletion trend in BSE, were removed at least partially to the Earth`s core during early differentiation events.

**Chalcophile elements in the BSE**. The large depletion of the chalcogens S, Se, and Te, which exceeds that of highly siderophile elements (HSE), led to the concept of a Hadean sulfide matte that greatly depleted the abundances of S, Se, Te, and the HSE Re, Os, Ir, Ru, Pt, Rh, Pd, and Au (e.g., O'Neill 1991, Ballhaus et al. 2017). It is possible that the abundances of the volatile chalcophile elements Pb, Cd, In, Tl, and Hg were modified by formation of the sulfide matte and/or by evaporative loss during accretion. Fish et al. (1960) suggested that a natural analog of Soxhlet extraction involving a S-bearing vapor, perhaps $S_2$, another sulfur allotrope, SiS (g), or another S-bearing gas could transport the chalcophile volatiles Bi, In, Pb, Tl. More recently, Heck (2022) and Steenstra et al. (2023, 2024) studied melt – gas partitioning of moderately volatile chalcophile elements (Cu, Zn, Ge, As, Ag, Cd, In, Pb). But if sulfide – gas partitioning is involved an open system is required, i.e. gravitational escape of the volatilized species, otherwise no depletions are produced (Fegley et al. 2020). Kiseeva and Wood (2013, 2015) studied sulfide – silicate melt partitioning of volatile elements with chalcophile and lithophile behavior (including Cd, In, Pb, Tl) and found the partition coefficients are sensitive to prevailing oxygen and sulfur fugacities. Chalcophile behavior of halogens occurs under some conditions, e.g., iodine in meteoritic troilite (Goles and Anders 1961, 1962, Goswami et al. 1998), the Cl-bearing mineral djerfisherite [$K_6(Cu,Ni,Fe)(Fe,Ni,Cu)_{24}S_{26}Cl$] in enstatite chondrites and Earth's mantle (Ebel and Sack 2013), partitioning of Cl, Br, and I into sulfide liquids (Lodders 1991, Steenstra et al. 2020). But the application of the melt – vapor and sulfide – silicate melt partitioning work to formation of a Hadean sulfide matte needs to be explored in more detail under appropriate (and possibly variable) fugacities ($O_2$, $S_2$, halogen), P, and T.

**BSE volatile element inventory**. Another question is the assessment of the volatile element budget of the BSE. If condensation (or evaporation) fractionation within a solar gas applies, there should not be a difference in relative abundances of lithophile, siderophile, and chalcophile elements of similar volatility within the *entire* Earth. However, we only have access to reliable estimates about the volatility-related fractionations of lithophile elements for the Earth's surface (atmosphere + hydrosphere + crust) and a part of the upper mantle, which are then used to give bulk silicate Earth abundances (e.g., see section 6.2 of Fegley et al. 2020). But even these data are uncertain to some extent with abundances of the most volatile elements generally being the most uncertain. The elements with the largest (≥40%, or unknown) uncertainties in Palme and O'Neill (2014) are B, F, S, Cl, Cu, Se, Br, Mo, Ag, Sb, Te, I, Cs, Hg, and Bi. With the exception of Mo these elements are moderately or highly volatile elements.

Finally, there is the conceptual issue of when the volatile element inventory of the Earth was established. Using solar condensation temperatures assumes that the building blocks of the Earth obtained their volatile complement by condensation (or evaporation) in equilibrium with the H and He-rich solar gas and not by evaporation from already accreted Earth materials after the moon-forming giant impact or other possible evaporation fractionations during accretion in the absence of H and He rich gas. Condensation (evaporation) temperatures of elements are significantly higher in anhydrous BSE vapor than in H and He-rich solar gas with the exception of some easily oxidized HSE (Ru, Os, Re) that can become much more volatile (Fegley et al. 2023).

**Nebular vs. planetary origin of volatile element depletions**. The origin of the volatile element depletion in meteorites and rocky planets is a long-standing question. Originally Anders (1964) proposed volatile element depletion resulted from variable mixtures of volatile-rich matrix and volatile-poor chondrules. Anders’ two component model is generally discounted as it is now clear chondrules are not poor in all volatiles (cf. Palme et al. 1988). Wasson and colleagues (Wasson and Chou 1974, Wai and Wasson 1977, Wasson 1977) attributed volatile element depletion in chondrites to incomplete condensation as the solar nebula cooled and dissipated. Using potassium stable isotopes, Tian et al. (2021) showed a correlation between volatile loss and planetary size for Earth, Mars, the Moon, and asteroid 4 Vesta, the parent body of EHD meteorites: larger bodies have higher gravity and less volatile loss. This is an attractive model that needs further testing with other MVE isotopes.

Wasson’s incomplete condensation model has stimulated the most theoretical work. Cassen (2001, Humayun and Cassen 2006) showed condensation of MVE into coagulating solids in a cooling solar nebula could reproduce MVE depletions in CO and CV, but not CM, carbonaceous chondrites. Cassen used a relatively quiescent nebula with only radiative heat transport that did not include convective heat and mass transport (Cassen 2001, 1993). Ciesla (2008) reexamined nebular models of MVE depletions and showed that the MVE depletion patterns in carbonaceous chondrites are only reproduced by a narrow range of physical parameters that are inconsistent with astronomical observations of protoplanetary disks around young stars. Ciesla’s models had more nebular turbulence than those of Cassen. Vollstaedt et al. (2020) involved photoevaporation of gas from the outer surface of the solar nebula to decrease nebular surface density (and hence opacity, temperature, and pressure) during MVE condensation into grains. They could match the MVE depletions in CM, CO, CV, and CK carbonaceous chondrites with reasonable physical parameters. Finally, Sengupta et al. (2022) used mass loss driven by disk winds to drive nebular cooling during MVE condensation. But none of the theoretical models explain why MVE depletions (volatility trends) are most evident in carbonaceous but not in ordinary or enstatite chondrites (cf. Figures 3-11).

# 6 Summary

The following major points are covered in this paper.
(1) Table 1 gives 50% condensation temperatures ($T_{50}$) for all naturally occurring elements and Pu at $10^{-2}$ to $10^{-8}$ bar total pressure for solar composition material.
(2) As shown in the Appendix, condensation temperatures are mainly controlled by the Gibbs energy of condensation reactions and also by the Gibbs energy of ideal mixing if elements (compounds) condense in a solution. The additional Gibbs energy change due to non-ideal solution, i.e., activity coefficients ≠1, is a secondary effect.
(3) The theoretically correct relationships between condensation temperature and fraction condensed ($\alpha_M$) is derived from mass balance and chemical thermodynamic considerations.

These relationships are log [$1/(1-\alpha_M)$] (for major elements) or log [$\alpha_M /(1-\alpha_M)$] (for elements in solid solutions) with *inverse* condensation temperatures ($K^{-1}$).
(4) The maximum amount of element condensed per $K^{-1}$, i.e., the maximum in [$d\alpha_M/d(1/T)$] is at the inflection point in the logistic (sigmoid) curve for an element, which is also at (or close to) the 50% condensation temperature.
(5) Plots of normalized elemental abundances versus 50% condensation temperatures (i.e., volatility trends) for meteorites and planets are qualitative indicators of elemental fractionations due to volatility. The popular volatility trend plots ($\alpha$ versus $T_{50}$) do not use the theoretically correct and quantitative relationship between condensation temperature and fraction condensed.
(6) Volatility trend plots for average elemental abundances in CM, CO, CV, CR, H, L, LL, EH, EL chondrites show different "trends" for moderately and highly volatile elements, which may be linear, curved, a step function, or plateau. A comparison of three abundance sets for CM and CV chondrites shows trends depend on which elements are plotted, which data sources are used, and which temperature range is considered.
(7) Proposed nebular mechanisms for the volatile element depletions in chondrites and the Earth are reviewed.
(8) Some possible implications of volatile element abundances in the BSE are reviewed and a preferred model for accretion of the Earth is presented.

Consideration of the volatility trends in chondrites and the bulk silicate Earth lead to the following questions:
Why do only carbonaceous chondrites show a clear volatility trend?

Why does the BSE apparently show a carbonaceous chondritic-like volatility trend if Zn and other volatile elements originated from non-carbonaceous chondritic material? Or do the trends look alike because terrestrial volatiles were sourced from carbonaceous chondritic material?

Further work is suggested to resolve two controversial issues:

The abundance of fluorine in enstatite chondrites, which looks anomalously high in Figures 10 and 11, is based mainly on analyses done in the 1960s (Table 2 in Lodders and Fegley 2023). Their discussion of the analytical data suggests it may be useful to remeasure F abundances in EH and EL chondrites using modern methods.

If one assumes a monotonic decrease, indium appears anomalous in the BSE volatility trend (see Fig. 13). This is not the case in the preferred model (Fig. 15) where Theia delivers indium. Witt-Eickschen et al. (2009) considered the possibility that indium is either too abundant in the BSE or erroneously too volatile. They propose the apparent problem is resolved if indium has about the same condensation temperature as zinc. This requires an activity coefficient of ca. 0.06 for InS dissolution in FeS at 697 K ($10^{-4}$ bar) where FeS is stable, but S is not yet 50% condensed. As mentioned in the Appendix, compounds such as $FeIn_2S_4$ indicate nonideal dissolution with activity coefficients <1 for InS in FeS. But the absence of activity coefficient data meant ideal solution was assumed to compute $T_{50}$ for indium. The InS activity coefficient in FeS can be measured by electromotive force measurements (e.g., see the discussion of emf measurements in Kubaschewski and Alcock 1979) and then used in condensation calculations.

## Acknowledgments

Work at Washington University is supported by NSF Astronomy Program Grant AST-2108172 and by the McDonnell Center for the Space Sciences. DE, BF, KL, and KM thank the International Space Science Institute for the invitation to the workshop. BF thanks JS Lewis for introducing him to "condensation calculations" in the 1970s, and BF and KL also thank Herbert Palme for discussions about chemical thermodynamics and nebular condensation over the past 40+ years.

## Appendix

**Interpolation of condensation temperatures**. Interpolation of 50% condensation temperatures at intermediate pressures in the range of $10^{-2}$ to $10^{-8}$ bar is done with linear fit equations:

$$\frac{10^4}{T_{50}} = A + BlogP$$

For example, the 50% condensation temperature equation for oxygen is

$$\frac{10^4}{T_{50}}(oxygen) = 40.12654 - 3.72440logP$$

If desired, the fits can be done with inverse temperature and then the A and B coefficients are correspondingly smaller.

**Thermodynamic data sources for pure compounds**. Thermodynamic data for over 2,000 compounds used in the calculations are listed in Fegley and Lodders (1994). Unless specified otherwise in the discussion, additional compounds and new (or revised) thermodynamic data are from the NIST-JANAF Tables (Chase 1998), the US Geological Survey tabulation of Robie and Hemingway (1995), Barin (1989), Knacke et al. (1991), and the IVTAN code thermodynamic database described by Belov et al (1999) and Gurvich et al. (1989, 1993, 1996). As discussed in Fegley et al. (2023), the IVTAN thermodynamic database is the continually updated Russian counterpart to the JANAF Tables. Vapor pressure data for calculating the 50% condensation temperatures of $H_2$ are from Honig and Hook (1960), from Hultgren et al. (1973a) for liquid He and Ne ice, from Table 1.20 of Lodders and Fegley (1998) for C (as $CH_4$ ice), and Ar as argon clathrate hydrate ($Ar \cdot 6H_2O$ ice), Marboeuf et al. (2012) for Kr clathrate hydrate, Fray et al. (2010) for Xe clathrate hydrate, and from Haudenschild (1970) for N (as $NH_3 \cdot H_2O$ ice). The 50% condensation temperatures for $H_2$ are fictive values at constant assumed total pressure, which drops to 43% of the initial value with half $H_2$ condensed (e.g., see Lewis 1972). Helium never condenses because condensation temperatures for liquid He are below the 3 K cosmic background temperature.

*Sodium and Potassium.* Sodium and potassium condense into feldspar, which is modeled as a solution of anorthite, albite, and potassium feldspar using activity coefficients for the albite and potassium feldspar components computed from Elkins and Grove (1990). Above 1000 K, the activity coefficient equations for albite (Na) and potassium feldspar (K) are

$$log\gamma_{Na} = 3.6 - \frac{3530}{T}$$

$$log\gamma_{K} = 6.3 - \frac{4800}{T}$$

Ideal solution is assumed below 1000 K.

*Rubidium*. Twenty-three Rb-bearing gases and 18 Rb-bearing solids are included in the calculations. In the vicinity of its 50% condensation temperature as $RbAlSi_3O_8$ dissolved in feldspar, RbCl is the major Rb-bearing gas closely followed by monatomic Rb.

Thermodynamic data for the two Rb aluminosilicates included in the calculations are all estimates. The Rb analog of leucite is considered first. Hirao and Soga (1982) measured the absolute entropies of $MAlSi_2O_6$ where M = K (leucite), Rb, and Cs (pollucite) but their $S°_{298}$ values for leucite and pollucite disagree with those in Robie and Hemingway (1995) and Ogorodova et al (2003) by large amounts and are incorrect. The $S°_{298}$ was estimated using ionic entropy values from Latimer (1951) and the assumption of $\Delta S_{298} = 0$ for the reaction

$$Rb^+ + KAlSi_2O_6 \text{ (leucite)} = K^+ + RbAlSi_2O_6 \text{ (Rb leucite)}$$

The standard enthalpy of formation at 298.15 K was estimated as the mean of the values for leucite and pollucite. The heat capacity for T = 298 K to 1100 K (limited by the melting point of $Rb_2SiO_3$ at 1143 K) was estimated using a Neumann – Kopp approach ($\Delta C_P = 0$) for the reaction:

$$\tfrac{1}{2} Rb_2SiO_3 \text{ (metasilicate)} + KAlSi_2O_6 \text{ (leucite)} = \tfrac{1}{2} K_2SiO_3 \text{ (metasilicate)} + RbAlSi_2O_6 \text{ (Rb leucite)}$$

The entropy and high temperature heat capacity of $RbAlSi_3O_8$ were estimated in a similar manner. The standard enthalpy of formation at 298.15 K was estimated from the reaction:

$$RbAlSi_2O_6 \text{ (Rb leucite)} + SiO_2 \text{ (quartz)} = RbAlSi_3O_8 \text{ (feldspar)} \quad \Delta H°_{298} = -21.5 \text{ kJ mol}^{-1}$$

This $\Delta H°_{298}$ value is in trend with the analogous reactions for Na (dehydrated analcime – albite $\Delta H°_{298} = -29.81$ kJ mol$^{-1}$) and K (leucite – sanidine $\Delta H°_{298} = -25.67$ kJ mol$^{-1}$), i.e., about 4 kJ mol$^{-1}$ smaller. For comparison at 1000 K, the estimated equilibrium constant (log $K_f$ value) from this work for formation of $RbAlSi_3O_8$ from the elements is 168.05 versus 167.95 calculated by Wood et al. (2019). Ideal solution 50% condensation temperatures at $10^{-4}$ bar for Rb are 1033 K (this work), ~1080 K (Wasson 1985 citing Grossman and Larimer 1974), and 800 K (Lodders 2003).

There are no data for $RbAlSi_3O_8$ activity coefficients in feldspar and an estimate ($\log \gamma = 130/T$) was used from comparison of two approaches: (1) activity coefficients for Rb salts in the RbF – KF ($\log \gamma = 130/T$) and RbCl – KCl ($\log \gamma = 80/T$) systems (Sangster and Pelton 1987), and (2) from Beswick's (1973) distribution coefficient (F) measurements of Rb/K ratios in sanidine and chloride vapor assuming $\log \gamma = \log (1/F) = 1146/T - 0.37$. At 1000 K, the estimated values for the Rb feldspar activity coefficient are 1.4 (RbF – KF), 1.2 (RbF – KCl), and 6.0 (Beswick), respectively.

*Cesium*. Twenty-five Cs-bearing gases and twenty Cs-bearing solids are included in the calculations. In the vicinity of its 50% condensation temperature as $CsAlSi_3O_8$ dissolved in feldspar, CsCl is the major Cs-bearing gas followed by monatomic Cs.

Thermodynamic data for pollucite are from Xu et al. (2001) and Ogorodova et al. (2003). The pollucite thermodynamic data give a Cs vapor pressure in agreement with the Knudsen effusion mass spectrometry measurements of Odoj and Hilpert (1980). Taylor et al. (1989) measured a slightly more negative (by 1%) standard enthalpy of formation from equilibria in aqueous solution at 200 °C. The low temperature adiabatic calorimetry of Bennington et al. (1983) gives a standard entropy <0.5% larger than that of Ogorodova et al. (2003). In contrast to pollucite, thermodynamic data for the Cs analog of feldspar $CsAlSi_3O_8$ had to be estimated (as also done by Wood et al. 2019 but using different methods). The $S°_{298}$ was estimated using ionic entropy values from Latimer (1951) and the assumption of $\Delta S_{298} = 0$ for the reaction

$$Cs^+ + KAlSi_3O_8 \text{ (sanidine)} = K^+ + CsAlSi_3O_8 \text{ (feldspar)}$$

The standard enthalpy of formation at 298.15 K was estimated from the reaction:

$$CsAlSi_2O_6 \text{ (pollucite)} + SiO_2 \text{ (quartz)} = CsAlSi_3O_8 \text{ (feldspar)} \quad \Delta H°_{298} = -21.5 \text{ kJ mol}^{-1}$$

The same $\Delta H°_{298}$ value was used for Rb and Cs feldspars for two reasons: (1) the $\Delta H°_f$ at 298 K for Rb leucite is an estimate and (2) trends in $\Delta H°_{298}$ values for formation of double oxides such as borates, carbonates, chromates, sulfates, and uranates from the constituent oxides show approximately the same values for Rb and Cs within a few kJ $mol^{-1}$. The heat capacity for T = 298 K to 1100 K (limited by the melting point of $Cs_2SiO_3$ at 1100 K) was estimated using a Neumann – Kopp approach ($\Delta C_P = 0$) for the reaction:

$$\tfrac{1}{2} Cs_2SiO_3 \text{ (Cs metasilicate)} + KAlSi_3O_8 \text{ (sanidine)} = \tfrac{1}{2} K_2SiO_3 \text{ (K metasilicate)} + CsAlSi_3O_8 \text{ (feldspar)}$$

The metasilicates are preferable to the oxides for two reasons: (1) the low melting point of $Cs_2O$ (768 K) limits the maximum temperature of the derived $C_P$ equation of Cs feldspar to 768 K, which is lower than the condensation temperature of the Cs feldspar at $10^{-2}$ and $10^{-4}$ bar total pressure and (2) the enthalpies of fusion and heat capacities for $Cs_2O$ and $K_2O$ are unknown and are estimates. For comparison this work gives an estimated equilibrium constant ($\log K_f$ value) for formation of $CsAlSi_3O_8$ from the elements of 168.25 versus 168.09 calculated by Wood et al. (2019) at 1000 K.

There are no data for $CsAlSi_3O_8$ activity coefficients in feldspar and an estimate (log $\gamma$ = 630/T) was used from comparison of three approaches: (1) activity coefficients for Cs salts in the CsCl – KCl (log $\gamma$ = 630/T), CsBr – KBr (log $\gamma$ = 470/T), CsI – KI (log $\gamma$ = 260/T) systems (Sangster and Pelton 1987), (2) from the $Cs_2SO_4$ activity coefficient (log $\gamma$ = 650/T) derived from its solubility limit in $K_2SO_4$ (Levin et al 1973), and (3) from Eugster's (1955) distribution coefficient (F) measurements of Cs/K ratios in sanidine and vapor assuming log $\gamma$ = log(1/F) = 1108/T – 0.58. At 1000 K, the estimated values for the Cs feldspar activity coefficient are 4.3 (CsCl – KCl), 3.0 (CsBr – KBr), 1.8 (CsI – KI), 4.5 ($Cs_2SO_4$ – $K_2SO_4$), and 3.4 (Eugster), respectively.

*Halogens*. Fegley and Schaefer (2010) and Lodders and Fegley (2023) describe halogen condensation at $10^{-4}$ bar total pressure. This work extends calculations to other total pressures. The major F- and Cl-bearing gases at the 50% condensation temperatures are HF and HCl. The major Br-bearing gas at the Br 50% condensation temperature is HBr, which comprises 96.5% ($10^{-8}$ bar) to 99.6% ($10^{-2}$ bar) of all gaseous bromine. Indium monobromide InBr comes second and includes almost all the rest of gaseous bromine (3.5% to 0.4%) over the same pressure range. The major I-bearing gas at the I 50% condensation temperature is HI, which is almost 100% ($10^{-2}$ bar) to 97.95% ($10^{-8}$ bar) of all gaseous iodine. With decreasing pressure, the second most abundant I-bearing gas switches from $SnI_2$ ($10^{-4}$% at $10^{-2}$ bar) to $GaI_3$ (0.004% at $10^{-4}$ bar to 2.0% at $10^{-8}$ bar).

The activity coefficients for NaBr and NaI dissolution in NaCl are from Sangster and Pelton (1987) and are given by the equations:

$$log\gamma_{NaBr} = \frac{287}{T} - 0.12$$

$$log\gamma_{NaI} = \frac{1{,}100}{T}$$

*Indium and Thallium*. Indium forms a thiospinel $FeIn_2S_4$ (known as indite) with iron sulfide (Mammedov et al. 2021). Thallium forms several ternary iron sulfides, e.g., $TlFeS_2$, $Tl_3Fe_2S_4$ (Zabel and Range 1979, Welz et al. 1989). Compound formation indicates indium and thallium sulfide dissolution in FeS is nonideal with activity coefficients less than unity. In the absence of any thermodynamic data ideal solution is assumed.

**Gibbs energy of mixing and activity coefficients**. Many groups have used activity coefficients to model nonideal solutions of minor (trace) elements in more abundant host phases. Larimer (1973) considered nonideal solution of Pb, Bi, and Tl in Fe metal and of InS and CdS in FeS. Kelly and Larimer (1977) computed 50% condensation temperatures for nonideal solution of Ga, Ge, and Pd in Fe alloy at high temperature. Wai and Wasson (1977, 1979) computed 50% condensation temperatures for nonideal solution of As, Au, Cu, Ga, Ge, and Sb in Fe alloy and ZnS in FeS. Sears (1978) considered several of the same elements. Boynton (1978) suggested Th, U, and Pu will have activity coefficients >1 in solid solution in perovskite. Davis and Grossman (1979) modeled ideal and nonideal condensation of REE oxides into perovskite and preferred ideal solution models to match observed REE patterns in Ca, Al-rich inclusions. Fegley (1980) calculated nonideal solution condensation of $BaTiO_3$ in perovskite from an assymetric solution Redlich – Kister analysis of the $BaTiO_3$ – $CaTiO_3$ phase diagram of DeVries and Roy (1955). At $5\times10^{-3}$ bar pressure along the Lewis (1974) nebular adiabat, $\gamma(BaTiO_3) = 10$ at 1500 K, the Ba 50% condensation temperature. Fegley and Lewis (1980) calculated nonideal solution condensation temperatures for P as $Fe_3P$ and in metal alloy and for Na and K in feldspar. Fraser and Rammensee (1982) measured activity coefficients in the Fe – Ni – Co system and used the METKON code (Palme and Wlotzka 1976) to compute alloy compositions in equilibrium with solar composition gas. Kornacki and Fegley (1986) computed the effect of variable activity coefficients on condensation of Ba, Sr, Pu, Th, and U dissolved as oxides in perovskite ($CaTiO_3$). Lodders (2003) calculated nonideal solution condensation temperatures for As, Au, Cu, Ga, Ge, and Sb dissolution in Fe alloy using the same activity coefficient data as prior workers (Sears 1978, Wai and Wasson 1977, 1979). Wood et al. (2019) repeated the earlier calculations and also considered nonideal solution of several additional elements in metal alloy, oxide, and silicate host phases. Many of their activity coefficients were estimated from crystal/melt distribution coefficients, the semi-empirical Miedema model for metal alloys, or from a lattice strain model used for oxides, silicates, and sulfides. Before proceeding it is worth quantitatively considering the effects of activity coefficients for the Gibbs energy of condensation reactions.

The relationship between the activity a, activity coefficient $\gamma$, and mole fraction X is

$$a = \gamma X$$

The activity coefficient is unity for an ideal solution and is either <1 or >1 for nonideal solutions. Negative deviations from ideality are indicated by e.g., compound formation, mean $\gamma$ <1 and higher condensation temperatures than ideal solution. Positive deviations from ideality are indicated by e.g., unmixing into two separate phases, mean $\gamma$ >1 and lower condensation temperature than ideal solution. The size of the activity coefficient correction term (excess Gibbs energy)

$$RTln\gamma = \Delta\bar{G}^E = \Delta\bar{H}^E - T\Delta\bar{S}^E$$

relative to the ΔG of condensation determines the importance of the activity coefficient term for the condensation temperature of a minor (trace) element (compound) dissolved in a host phase. For example, considering reaction (1) for M (gas) = M(alloy) we can write

$$\Delta G_1 = \Delta H_1 - T\Delta S_1$$

Where the enthalpy and entropy of reaction are the same as previously defined in Equation (11). With apologies to thermodynamic notation purists, we use simplified notation that is easier for most readers to understand. All quantities are on a molar basis. We can also write

$$\Delta G_1 = \Delta G^o_{pure} + \Delta \bar{G}^I + \Delta \bar{G}^E$$

The Gibbs energy of condensation for reaction (1) is the sum of three Gibbs energy terms: that for the pure element (or compound) M, the ideal (I) solution of mixing term, and the non-ideal (E, excess) solution of mixing term. (The degree sign superscript denotes the pure element (or compound) at one bar pressure and is the value listed in the JANAF Tables or another thermodynamic data compilation. The overbars denote partial molal quantities, e.g., for element M in an alloy.) In general (e.g., see arguments by Ebel 2005, Ebel and Grossman 2005, Fegley et al. 2023 for silicate melts),

$$\Delta G_1 > \Delta \bar{G}^E$$

The ideal solution of mixing term is the partial molal ideal solution of mixing and is defined as

$$\Delta \bar{G}^I = RT ln X_M$$

Taking Fe – a major element – as an example, at $10^{-4}$ bar total pressure, pure Fe metal is 50% condensed at 1331 K and Fe metal in ideal solution in metal alloy is 50% condensed at 1334 K, only 3 K higher. At 1334 K, $\Delta G°_{pure}$ = −217,100 J/mol for the reaction

Fe (gas) = Fe (metal)

The partial molal ideal solution of mixing of Fe in the metal alloy (with $X_{Fe}$ = 0.913) is $\Delta \bar{G}^I$ = −1,010 J/mol for dissolution of pure Fe metal into the alloy

Fe (metal) = Fe (alloy)

and $RT ln \gamma_{Fe}$ = −11 J/mol from Fraser and Rammensee (1982) to account for

Fe (ideal alloy) = Fe (nonideal alloy)

(For reference, the other constituents of the alloy are primarily Ni and Co with mole fractions of $X_{Ni}$ = 0.083 and $X_{Co}$ = 0.00396.) This comparison shows that the ideal solution term is an insignificant part (0.5%) of the total Gibbs energy term. This result applies generally to chalcophile, lithophile, and siderophile elements (and compounds) that are major components of a solution, not only to iron metal. The 50% condensation temperature of a pure element (compound) is not greatly different from that of the 50% condensation temperature for the element (compound) that is a major component dissolved in an ideal solution. However, this is not the case for minor or trace elements dissolved in a host phase, which are considered next.

Palladium condensation illustrates the effect of ideal mixing and an activity coefficient <1 on condensation temperatures of a minor (trace) element dissolved in solution. Pure Pd metal is 50% condensed at 901 K at $10^{-4}$ bar total pressure. The 50% condensation temperature increases to 1321 K for ideal solution of Pd in Fe,Ni-rich metal alloy. Nonideal solution using the "most likely" $\gamma$ = 0.4 increases the Pd condensation temperature to 1342 K (Palme et al. 1988). The Pd activity coefficient used by Palme et al. (1988) is the infinite dilution value for $\gamma$-Fe – Pd metal alloy from Hultgren et al. (1973). At 1342 K, $\Delta G^{\circ}_{pure}$ = −209,720 J/mol, the ideal solution term is $RT\ln X_{Pd}$ = −141,900 J/mol, and the activity coefficient term $RT\ln\gamma$ = −10,220 J/mol. The activity coefficient term is 2.8% of the total Gibbs energy term (−361,840 J/mol).

Thoria ($ThO_2$) condensation in a host phase such as hibonite or perovskite exemplifies the effect of an activity coefficient >1 on condensation temperatures. The results of Grossman (1973) and Kornacki and Fegley (1986) are used for illustration. Grossman (1973) calculated pure $ThO_2$ to condense at 1496 K at $10^{-3}$ bar total pressure. CONDOR code calculations confirm that value, give a 50% condensation temperature of 1478 K, and show $ThO_2$ gas is 99.99% of total gaseous thorium. The condensation reaction is

$$ThO_2 \text{ (gas)} = ThO_2 \text{ (thoria)}$$

Kornacki and Fegley (1986) computed that ideal solution of $ThO_2$ in perovskite increases the $ThO_2$ 50% condensation temperature to 1676 K at $10^{-3}$ bar, where perovskite first appears. Using $ThO_2$ activity coefficients >1, as argued by Boynton (1978), they computed lower 50% condensation temperatures (their Table 4). Taking $\gamma$ = 50 for this example, $ThO_2$ is 50% condensed at 1675.4 K. The ΔG values for the condensation reaction of pure thoria, ideal solution of $ThO_2$ in perovskite, and $RT\ln\gamma$ for $ThO_2$ are below:

$$\Delta G_1 = \Delta G^{o}_{pure} + RTX_{ThO_2} + RTln\gamma_{ThO_2}$$

$$\Delta G_1 = -436{,}830 - 152{,}110 + 54{,}490$$

The activity coefficient term is about 10% of the total Gibbs energy (−534,450 J/mol) and opposite in sign to the other two terms. The positive $RT\ln\gamma$ decreases the 50% condensation temperature but the effect is tiny for $ThO_2$ because it is so refractory with a large negative ΔG of condensation. Uranium dioxide condensation is another example, and, in this case, the same size activity coefficient gives a larger decrease in the 50% condensation temperature because $UO_2$ is less refractory than thoria (Kornacki and Fegley 1986).

## Declarations

Competing Interests: The authors declare that they have no conflict of interest.

## References

F. Albarède, Volatile accretion history of the terrestrial planets and dynamic implications. Nature **461**, 1227-1233 (2009)

C.M.O'D. Alexander, Quantitative models for the elemental and isotopic fractionations in chondrites: The non-carbonaceous chondrites. Geochim. Cosmochim. Acta **254**, 246-276 (2019a)

C.M.O’D. Alexander, Quantitative models for the elemental and isotopic fractionations in chondrites: The carbonaceous chondrites. Geochim. Cosmochim. Acta **254**, 277-309 (2019b)

E. Anders, Origin, age, and composition of meteorites. Space Sci. Rev. 3, 583-714 (1964)

E. Anders, On the depletion of moderately volatile elements in chondrites. Meteoritics **10**, 283-286 (1975)

E. Anders, Critique of “Nebular condensation of moderately volatile elements and their abundances in ordinary chondrites” by Chien M. Wai and John T. Wasson. Earth Planet. Sci. Lett. **36**, 14-20 (1977)

G. Avice, B. Marty, The iodine – plutonium – xenon age of the Moon – Earth system revisited. Philos. Trans. Roy. Soc. London **A372**, 20130260 (2014)

C. Ballhaus, R.O.C. Fonseca, C. Münker, A. Rohrbach, T. Nagel, I.M. Speelmans, H.M. Helmy, A. Zirner, A.K. Vogel, A. Heuser, The great sulfur depletion of Earth’s mantle is not a signature of mantle – core equilibration. Contrib. Mineral. Petrol. 172, 68 10pp (2017)

S.S. Barshay, Combined Condensation – Accretion Models of the Terrestrial Planets. PhD thesis, MIT, Cambridge MA (1981)

K.O. Bennington, R.P. Beyer, G.K. Johnson, Thermodynamic Properties of Pollucite (A Cesium – Aluminum Silicate). U.S. Bureau of Mines Report of Investigations 8779, 29pp (1983)

A.E. Beswick, An experimental study of alkali metal distributions in feldspars and micas. Geochim. Cosmochim. Acta **37**, 183-208 (1973)

W.V. Boynton, Fractionation in the solar nebula: condensation of yttrium and the rare earth elements. Geochim. Cosmochim. Acta **39**, 569-584 (1975)

W.V. Boynton, Fractionation in the solar nebula, II. Condensation of Th, U, Pu and Cm. Earth Planet. Sci. Lett. **40**, 63-70 (1978)

N. Braukmüller, F. Wombacher, C. Funk, C. Münker, Earth’s volatile element depletion pattern inherited from a carbonaceous chondrite-like source. Nat. Geosci. **12**, 564–569 (2019)

A.G.W. Cameron, The first ten million years in the solar nebula. Meteorit. Planet. Sci. **30**, 133-161 (1995)

A.G.W. Cameron, M.B. Fegley, Nucleation and condensation in the primitive solar nebula. Icarus **52**, 1-13 (1982)

P. Cassen, Why convective heat transport in the solar nebula was inefficient. Lunar Planet. Sci. **XXIV**, 261-262 (1993)

P. Cassen, Nebular thermal evolution and the properties of primitive planetary materials. Meteorit. Planet. Sci. **36**, 671-700 (2001)

F.J. Ciesla, Radial transport in the solar nebula: Implications for moderately volatile element depletions in chondritic meteorites. Meteorit. Planet. Sci. 43, 639-655 (2008)

N. Dauphas, The isotopic nature of the Earth’s accreting material through time. Nature **541**, 521-524 (2017)

N. Dauphas, C. Burkhardt, P.H. Warren, F.-Z. Teng, Geochemical arguments for an Earth-like Moon-forming impactor. Phil. Trans. R. Soc. A **372**: 20130244 (2014)

N. Dauphas, A. Pourmand, Thulium anomalies and rare earth element patterns in meteorites and Earth: Nebular fractionation and the nugget effect. Geochim. Cosmochim. Acta **163**, 234-261 (2015)

A.M. Davis, L. Grossman, Condensation and fractionation of rare earths in the solar nebula. Geochim. Cosmochim. Acta **43**, 1611-1632 (1979)

R.C. DeVries, R. Roy, Phase equilibria in the system $BaTiO_3$ – $CaTiO_3$. J. Am. Ceram. Soc. **38**, 142-146 (1955)

D.S. Ebel, Model evaporation of FeO-bearing liquids: Applications to chondrules. Geochim. Cosmochim. Acta **69**, 3183-3193 (2005)

D.S. Ebel, Condensation of rocky material in astrophysical environments. In Meteorites and the Early Solar System II, pp. 253-277, Eds. D. Lauretta, H.Y. McSween, Jr., University of Arizona Press (2006)

D.S. Ebel, C.M.O'D. Alexander, G. Libourel, Vapor-melt exchange – Constraints on chondrite formation conditions and processes. Chapter 6 in Chondrules: Records of Protoplanetary Disk Processes, eds. S. Russell, H.C. Connolly Jr. and A.N. Krot. Cambridge University Press, pp 151-174 (2018)

D.S. Ebel, L. Grossman, Condensation in dust-enriched systems. Geochim. Cosmochim. Acta **64**, 339-366 (2000)

D.S. Ebel, L. Grossman, Spinel-bearing spherules condensed from the Chicxulub impact-vapor plume. Geology **33**, 293-296 (2005)

D.S. Ebel, R.O. Sack, Djerfisherite: nebular source of refractory potassium. Contrib. Mineral. Petrol. **166**, 923-934 (2013)

D.S. Ebel, M.K. Weisberg, J.R. Beckett, Thermochemical stability of low-iron, manganese-enriched olivine in astrophysical environments. Meteorit. Planet. Sci. 47, 585-593 (2012)

L.T. Elkins, T.L. Grove, Ternary feldspar experiments and thermodynamic models. Am. Min. **75**, 544-559 (1990)

H.J.T. Ellingham, Reducibility of oxides and sulphides in metallurgical processes. J. Soc. Chem. Ind. **63**, 125-133 (1944)

H.P. Eugster, The cesium – potassium equilibrium in the system sanidine – water, Carnegie Inst. Washington Yearbook No. 54 (1954-1955), pp. 112-113 (1955)

M.B. Fegley, Jr., Chemical Fractionations in Solar Composition Material, PhD thesis, MIT, 168pp, https://dspace.mit.edu/handle/1721.1/52853

B. Fegley, Jr., Primordial retention of nitrogen by terrestrial planets and meteorites. Proc. 13$^{th}$ Lunar Planet. Sci. Conf. J. Geophys. Res. **88**, A853-A868 (1983)

B. Fegley, Jr., Kinetics of gas-grain reactions in the solar nebula. Space Sci. Rev. **92**, 177-200 (2000)

B. Fegley, Jr., A.S. Kornacki, The origin and mineral chemistry of Group II inclusions in carbonaceous chondrites. Lunar Planet. Sci. **XV**, 262-263 (1984)

B. Fegley, Jr., J.S. Lewis, Volatile element chemistry in the solar nebula: Na, K, F, Cl, and P. Icarus **41**, 439-455 (1980)

B. Fegley, Jr., K. Lodders, Chemical models of the deep atmospheres of Jupiter and Saturn. Icarus **110**, 117-154 (1994)

B. Fegley, Jr., K. Lodders, N.S. Jacobson, Volatile element chemistry during accretion of the Earth. Geochemistry **80**, 125594, 40pp (2020)

B. Fegley, Jr., K. Lodders, N.S. Jacobson, Chemical equilibrium calculations for bulk silicate Earth material at high temperatures. Geochemistry 83, 125961 35pp (2023)

B. Fegley, Jr., H. Palme, Evidence for oxidizing conditions in the solar nebula from Mo and W depletions in refractory inclusions in carbonaceous chondrites. Earth Planet. Sci. Lett. 72, 311-326 (1985)

B. Fegley, Jr., R.G. Prinn, Solar nebula chemistry: Implications for volatiles in the solar system. In The Formation and Evolution of Planetary Systems, eds. H.A. Weaver, L. Danly, pp. 171-211, Cambridge University Press, Cambridge UK (1989)

B. Fegley, Jr., L. Schaefer, Cosmochemistry. In Principles and Perspectives in Cosmochemistry, pp. 347-377, Springer Verlag, Berlin (2010)

R.A. Fish, G.G. Goles, E. Anders, The record in the meteorites III. On the development of meteorites in asteroidal bodies. Astrophys. J. 132, 243-258 (1960)

M. Fischer-Gödde, T. Kleine, Ruthenium isotopic evidence for an inner Solar System origin of the late veneer. Nature **541**, 525-527 (2017)

D.G. Fraser, W. Rammensee, Activity measurements by Knudsen cell mass spectrometry – the system Fe-Co-Ni and implications for condensation processes in the solar nebula. Geochim. Cosmochim. Acta **46**, 549-556 (1982)

N. Fray, U. Marboeuf, O. Brissaud, B.Schmitt, Equilibrium data of methane, carbon dioxide, and xenon clathrate hydrates below the freezing point of water. Applications to astrophysical environments. J. Chem. Eng. Data **55**, 5101-5108 (2010)

P.W. Gast, The isotopic composition of strontium and the age of stone meteorites – I. Geochim. Cosmochim. Acta 26, 927-943 (1962)

V. M. Goldschmidt, Geochemische Verteilungsgesetze der Elemente. Skrifter Norske Videnskaps-Akademi Oslo, I. Mat.-Naturv. Klasse No. 3, 3-17 (1923)

G.G. Goles, Sodium. In Handbook of Elemental Abundances in Meteorites. B. Mason, ed., pp. 109-113, Gordon & Breach, New York (1971a)

G.G. Goles, Potassium. In Handbook of Elemental Abundances in Meteorites. B. Mason, ed., pp. 149-169, Gordon & Breach, New York (1971b)

G.G. Goles, E. Anders, On the geochemical character of iodine in meteorites. J. Geophys. Res. **66**, 3075-3077 (1961)

G.G. Goles, E. Anders, Abundances of iodine tellurium and uranium in meteorites. Geochim. Cosmochim. Acta **26**, 723-737 (1962)

J.N. Goswami, S. Sahijpal, K. Kehm, C.M. Hohenberg, T. Swindle, J.N. Grossman, In situ determination of iodine content and iodine-xenon systematics in silicates and troilite phases in chondrules from the LL3 chondrite Semarkona. Meteorit. Planet. Sci. **33**, 527-534 (1998)

L. Grossman, Condensation in the primitive solar nebula. Geochim. Cosmochim. Acta **36**, 597-619 (1972)

L. Grossman, Refractory trace elements in Ca-Al-rich inclusions in the Allende meteorite. Geochim. Cosmochim. Acta **37**, 1119-1140 (1973)

L. Grossman, J.W. Larimer, Early chemical history of the solar system. Rev. Geophys. Space Phys. **12**, 71-101 (1974)

L. Grossman, E. Olsen, Origin of the high-temperature fraction of C2 chondrites. Geochim. Cosmochim. Acta **38**, 173-187 (1974)

A.N. Halliday, Mixing, volatile loss and compositional change during impact-driven accretion of the Earth. Nature. **427**, 505-509 (2004)

C. Haudenschild, Multi-phase ammonia-water system (Rev 1). JPL Space Programs Summary 37-66 vol III, pp. 4-9 (1970)

W. Heck, Experimental Degassing of Moderately Volatile Chalcophile Elements from Silicate Melt. MS Thesis, Tulane University (2022)

K. Hirao, N. Soga, The heat capacity and a phase transition in leucite-type compounds. Yogyo Kyokai Shi **90**, 390-396 (1982)

A. Holzheid, H.St.C. O'Neill, The Cr – $Cr_2O_3$ oxygen buffer and the free energy of formation of $Cr_2O_3$ from high-temperature electrochemical measurements. Geochim. Cosmochim. Acta 59, 475-479 (1995)

R.E. Honig, H.O. Hook, Vapor pressure data for some common gases. RCA Review 21, 360-368 (1960)

G.B. Hudson, B.M. Kennedy, F.A. Podosek, C.M. Hohenberg, The early solar system abundance of $^{244}Pu$ as inferred from the St. Severin chondrite. Proc 19th Lunar Planet. Sci. Conf. **19**, 547-557 (1989)

R. Hultgren, P.D. Desai, D.T. Hawkins, M. Gleiser, K.K. Kelley, D.D. Wagman, Selected Values of the Thermodynamic Properties of the Elements. American Society for Metals (1973a)

R. Hultgren, P.D. Desai, D.T. Hawkins, M. Gleiser, K.K. Kelley, Selected Values of the Thermodynamic Properties of Binary Alloys. American Society for Metals (1973b)

M. Humayun, P. Cassen, Processes determining the volatile abundances of the meteorites and terrestrial planets. In Origin of the Earth and Moon, pp. 3-23 (2006)

T.R. Ireland, B. Fegley, Jr., The solar system's earliest chemistry: Systematics of refractory inclusions. Intl. Geol. Rev. 42, 865-894 (2000)

J.S. Kargel, J.S. Lewis, The composition and early evolution of Earth. Icarus **105**, 1–25 (1993)

H.H. Kellogg, Vaporization chemistry in extractive metallurgy. Trans. Met. Soc. AIME **236**, 602-615 (1966)

W.R. Kelly, J.W. Larimer, Chemical fractionations in meteorites – VIII. Iron meteorites and the cosmochemical history of the metal phase. Geochim. Cosmochim. Acta **41**, 93-111 (1977)

E.S. Kiseeva, B.J. Wood, A simple model for chalcophile element partitioning between sulphide and silicate liquids with geochemical implications. Earth Planet. Sci. Lett. **383**, 68-81 (2013)

E.S. Kiseeva, B.J. Wood, The effects of composition and temperature on chalcophile and lithophile element partitioning into magmatic sulphides. Earth Planet. Sci. Lett. **424**, 280-292 (2015)

A.S. Kornacki, B. Fegley, Jr., The abundance and relative volatility of refractory trace elements in Allende Ca,Al-rich inclusions: implications for chemical and physical processes in the solar nebula. Earth Planet. Sci. Lett. 79, 217-234 (1986)

U. Krahenbühl, J.W. Morgan, R. Ganapathy, E. Anders, Abundance of 17 trace elements in carbonaceous chondrites. Geochim. Cosmochim. Acta **37**, 1353-1370 (1973)

O. Kubaschewski, C.B. Alcock, Metallurgical Thermochemistry. 5th ed., Pergamon Press (1979)

J.W. Larimer, Chemical fractionations in meteorites – I. Condensation of the elements. Geochim. Cosmochim. Acta **31**, 1215-1238 (1967)

J.W. Larimer, Chemical fractionations in meteorites – VII. Cosmothermometry and cosmobarometry. Geochim. Cosmochim. Acta **37**, 1603-1623 (1973)

J.W. Larimer, The cosmochemical classification of the elements, in Meteorites and the Early Solar System, eds. J.F. Kerridge, M.S. Matthews, pp. 375-389 (1988)

J.W. Larimer, E. Anders, Chemical fractionations in meteorites – II. Abundance patterns and their interpretation. Geochim. Cosmochim. Acta **31**, 1239-1270 (1967)

J.W. Larimer, J.T. Wasson, Refractory lithophile elements, in Meteorites and the Early Solar System, eds. J.F. Kerridge, M.S. Matthews, pp. 390-415 (1988)

W.M. Latimer, Methods of estimating the entropies of solid compounds. J. Amer. Chem. Soc. **73**, 1480-1482 (1951)

D.S. Lauretta, K. Lodders, The cosmochemical behavior of beryllium and boron. Earth Planet. Sci. Lett. **146**, 315-327 (1997)

M.R. Lee, C.L. Smith, S.H. Gordon, M.E. Hodson, Laboratory simulation of terrestrial meteorite weathering using the Bensour (LL6) ordinary chondrite. Meteorit. Planet. Sci. 41, 1123-1138 (2006)

E.M. Levin, J.T. Benedict, J.P. Sciarello, S. Monsour, The system $K_2SO_4$ – $Cs_2SO_4$. J. Am. Ceram. Soc. **56**, 427-430 (1973)

J.S. Lewis, Metal/silicate fractionation in the solar system. Earth Planet. Sci. Lett. 15, 286-290 (1972a)

J.S. Lewis, Low temperature condensation from the solar nebula. Icarus 16, 241-252 (1972b)

J.S. Lewis, The temperature gradient in the solar nebula. Science **186**, 440-443 (1974)

J.S. Lewis, Origin and composition of mercury, in Mercury, eds. F. Vilas, C.R. Chapman, M.S. Matthews, pp. 651-666, Univ. of Arizona Press, Tucson (1988)

J.S. Lewis, S.S. Barshay, B. Noyes, Primordial retention of carbon by the terrestrial planets. Icarus **37**, 190-206 (1979)

J.S. Lewis, R.G. Prinn, Kinetic inhibition of CO and $N_2$ reduction in the solar nebula. Astrophys. J. **238**, 357-364 (1980)

J.S. Lewis, R.G. Prinn, Planets and Their Atmospheres: Origin and Evolution. Academic Press, San Diego (1984)

C. Liebske, A. Khan, On the principal building blocks of Earth and Mars. Icarus **322**, 121-134 (2019)

M.E. Lipschutz, D.S. Woolum, Highly labile elements, Meteorites and the Early Solar System, eds. J.F. Kerridge, M.S. Matthews, pp. 462-487 (1988)

K. Lodders, Spurenelementverteilung zwischen Sulfid and Silikatschmelze und Kosmochemische Anwendungen, Ph.D. thesis, Univ. Mainz, Germany, 176 pp (1991)

K. Lodders, An oxygen isotope mixing model for the accretion and composition of rocky planets. Space Sci. Rev. **92**, 341-354 (2000)

K. Lodders, Solar system abundances and condensation temperatures of the elements. Astrophys. J. **591**, 1220-1247 (2003)

K. Lodders, Solar elemental abundances. In The Oxford Research Encyclopedia of Planetary Science, Oxford University Press (2020)

K. Lodders, Relative atomic solar system abundances, mass fractions, and atomic masses of the elements and their isotopes, composition of the solar photosphere, and compositions of the major chondrite groups. Space Sci. Rev. **217**:44 (33pp) (2021)

K. Lodders, B. Fegley, Jr., Lanthanide and actinide chemistry at high C/O ratios in the solar nebula. Earth Planet. Sci. Lett. **117**, 125-145 (1993)

K. Lodders, B. Fegley, Jr., An oxygen isotope model for the composition of Mars. Icarus **126**, 373-394 (1997)

K. Lodders, B. Fegley, Jr., Condensation chemistry of carbon stars. In Astrophysical Implications of the Laboratory Study of Presolar Materials, eds. T.J. Bernatowicz, E.K. Zinner, pp. 391-423, AIP Press (1997)

K. Lodders, B. Fegley, Jr., The Planetary Scientist's Companion, Oxford University Press (1998)

K. Lodders, B. Fegley, Jr., Solar system abundances and condensation temperatures of the halogens fluorine, chlorine, bromine, and iodine. Geochemistry **83**, 125957, 27pp (2023)

H.C.Lord, III, Molecular equilibria and condensation in a solar nebula and cool stellar atmospheres. Icarus **4**, 279-288 (1965)

A. Maltese, K. Mezger, The Pb isotope evolution of Bulk Silicate Earth: Constraints from its accretion and early differentiation history. Geochim. Cosmochim. Acta **271**, 179-193 (2020)

F.M. Mammedov, D.M. Babanly, I.R. Amiraslanov, D.B. Tagiev, M.B. Babanly, FeS – $Ga_2S_3$ – $In_2S_3$ system. Russ. J. Inorg. Chem. **66**, 1533-1543 (2021)

U. Marboeuf, N. Fray, O. Brissaud, B. Schmitt, D. Bockelée-Morvan, D. Gautier, Equilibrium pressure of ethane, acetylene, and krypton clathrate hydrates below the freezing point of water. J. Chem. Eng. Data **57**, 3408-3415 (2012)

B. Mason, Data of Geochemistry chapter B. Cosmochemistry, Part 1. Meteorites. U.S. Geological Survey Professional Paper 440-B-1, pp. B1-B132 (1979)

W.F. McDonough, S.-S. Sun, The composition of the Earth. Chem. Geol. **120**, 223–253 (1995)

K. Mezger, A. Maltese, H. Vollstaedt, Accretion and differentiation of early planetary bodies as recorded in the composition of the silicate Earth. Icarus **365**, 114497 12pp (2021)

K. Mezger, M. Schönbächler, A. Bouvier, Accretion of the Earth—Missing Components? Space Sci. Rev. **216**, 1-24 (2020)

R. Odoj, K. Hilpert, Mass spectrometric study of the evaporation of crystalline compounds in the $Cs_2O$ – $Al_2O_3$ – $SiO_2$ system. 1. The synthetic compound $CsAlSi_2O_6$ and the mineral pollucite. High Temperatures – High Pressures **12**, 93-98 (1980)

L.P. Ogorodova, L.V. Melchakova, I.A. Kiseleva, I.A. Belitsky, Thermochemical study of natural pollucite. Thermochim. Acta **403**, 251-256 (2003)

H.St.C. O'Neill, The origin of the Moon and the early history of the Earth – A chemical model. Part 2: the Earth. Geochim. Cosmochim. Acta 55, 1159-1172 (1991)

H. Palme, J.W. Larimer, M.E. Lipschutz, Moderately volatile elements, Meteorites and the Early Solar System, eds. J.F. Kerridge, M.S. Matthews, pp. 436-461 (1988)

H. Palme, H.St.C. O'Neill, Cosmochemical estimates of mantle composition. Treatise on Geochemistry, 2nd ed, Elsevier, Amsterdam, pp. 1–39 (2014)

H. Palme, F. Wlotzka, A metal particle from a Ca, Al-rich inclusion from the meteorite Allende, and the condensation of refractory siderophile elements. Earth Planet. Sci. Lett. 33, 45-60 (1976)

A.E. Ringwood, The Origin of the Earth and Moon, Springer Verlag, New York (1979)

A.E. Ringwood, S.E. Kesson, Basaltic magmatism and the bulk composition of the Moon. II. Siderophile and volatile elements in Moon, Earth and chondrites: implications for lunar origin. Moon **16**, 425–464 (1977)

S.P. Ruden, D.N.C. Lin, The global evolution of the primordial solar nebula. Astrophys J. **308**, 883-901 (1986)

J. Sangster, A.D. Pelton, Phase diagrams and thermodynamic properties of the 70 binary alkali halide systems having common ions. J. Phys. Chem. Ref. Data 16, 509-561 (1987)

A.R. Sarafian, S.G. Nielsen, H.R. Marschall, G.A. Gaetani, E.H. Hauri, K. Righter, E. Sarafian, Angrite meteorites record the onset and flux of water to the inner solar system. Geochim. Cosmochim. Acta **212**, 156-166 (2017)

S.K. Saxena, G. Eriksson, Low- to medium-temperature phase equilibria in a gas of solar composition. Earth Planet. Sci. Lett. **65**, 7-16 (1983a)

S.K. Saxena, G. Eriksson, High temperature phase equilibria in a solar-composition gas. Geochim. Cosmochim. Acta **47**, 1865-1874 (1983b)

M. Schönbächler, R. Carlson, M. Horan, T. Mock, E.H. Hauri, Heterogeneous accretion and the moderately volatile element budget of the Earth. Science **328**, 884-887 (2010)

D.W. Sears, Condensation and the composition of iron meteorites. Earth Planet. Sci. Lett. **41**, 128-138 (1978)

D.W. Sears, Formation of E chondrites and aubrites – A thermodynamic model. Icarus **43**, 184-202 (1980)

D. Sengupta, P.R. Estrada, J.N. Cuzzi, M. Humayun, Depletion of moderately volatile elements by open-system loss in the early solar nebula. Astrophys. J. **932**:82 (28pp) (2022)

P.A. Sossi, I.L. Stotz, S.A. Jacobson, A. Morbidelli, H.St.C. O’Neill, Stochastic accretion of the Earth. Nature Astronomy **6**, 951-960 (2022)

E.S. Steenstra, C.J. Renggli, J. Berndt, S. Klemme, Evaporation of moderately volatile elements from metal and sulfide melts: Implications for volatile element abundances in magmatic iron meteorites. Earth Planet. Sci. Lett. 622, 118406 13pp (2023)

E.S. Steenstra, C.J. Renggli, J. Berndt, S. Klemme, Quantification of evaporative loss of volatile metals from planetary cores and metal-rich planetesimals. Geochim. Cosmochim. Acta 384, 93-110 (2024)

E.S. Steenstra, F. van Haaster, R. van Mulligan, S. Flemetakis, J. Berndt, S. Klemme, W. van Westrenen, An experimental assessment of the chalcophile behavior of F, Cl, Br and I: Implications for the fate of halogens during planetary accretion and the formation of magmatic ore deposits. Geochim. Cosmochim. Acta 273, 275-290 (2020)

A. Stracke, H. Palme, M. Gellissen, C. Münker, T. Kleine, K. Birbaum, D. Günther, B. Bourdon, J. Zipfel, Refractory element fractionation in the Allende meteorite: Implications for solar nebula condensation and the chondritic composition of parent bodies. Geochim. Cosmochim. Acta **85**, 114-141 (2012)

S.-S. Sun, Chemical composition and origin of the earth’s primitive mantle. Geochim. Cosmochim. Acta **46**, 179–192 (1982)

R.B. Symonds, W.I. Rose, G.J.S. Bluth, T.M. Gerlach, Volcanic-gas studies: Methods, results, and applications. In Volatiles in Magmas, eds. M.R. Carroll, J.R. Holloway, pp. 1-66, Mineralogical Society of America, Washington DC (1994)

H. Takahashi, J. Gros, H. Higuchi, J.W. Morgan, E. Anders, Volatile elements in chondrites: metamorphism or nebular fractionation? Geochim. Cosmochim. Acta **42**, 1859-1869 (1978)

T. Tanaka, A. Masuda, Rare-earth elements in matrix, inclusions, and chondrules of the Allende meteorite. Icarus **19**, 523-530 (1973)

P. Taylor, S.D. DeVaal, D.G. Owen, Stability relationships between solid cesium aluminosilicates in aqueous solutions at 200 °C. Canad. J. Chem. **67**, 76-81 (1989)

Z. Tian, T. Magna, J.M.D. Day, K. Mezger, E.E. Scherer, K. Lodders, R.C. Hin, P. Koefoed, H. Bloom, K. Wang, Potassium isotope composition of Mars reveals a mechanism of planetary volatile retention. PNAS 118, e2101155118 7pp (2021)

H.C. Urey, The Planets: Their Origin and Development. Yale University Press (1952)

H.C. Urey, On the dissipation of gas and volatilized elements from protoplanets. Astrophys. J. Suppl. **1**, 147-173 (1954)

H. Vollstaedt, K. Mezger, Y. Allibert, Carbonaceous chondrites and the condensation of elements from the solar nebula. Astrophys. J. **897**:82 (14 pp) (2020)

C.M. Wai, J.T. Wasson, Nebular condensation of moderately volatile elements and their abundances in ordinary chondrites. Earth Planet. Sci. Lett. **36**, 1-13 (1977)

C.M. Wai, J.T. Wasson, Nebular condensation of Ga, Ge and Sb and the chemical classification of iron meteorites. Nature **282**, 790-793 (1979)

H.S. Wang, C.H. Lineweaver, T.R. Ireland, The volatility trend of protosolar and terrestrial elemental abundances. Icarus **328**, 287–305 (2019)

H. Wänke, Constitution of terrestrial planets. Phil. Trans. Roy. Soc. London **303A**, 287-302 (1981)

H. Wänke, H. Baddenhausen, H. Palme, B. Spettel, On the chemistry of the Allende inclusions and their origin as high temperature condensates. Earth Planet. Sci. Lett. **23**, 1-7 (1974)

H. Wänke G. Dreibus, Chemical composition and accretion history of terrestrial Planets. Phil. Trans. Roy. Soc. London. Series A, **325**, 545-557 (1988)

H. Wänke, G. Dreibus, E. Jagoutz, Mantle chemistry and accretion history of the Earth, in Archean Geochemistry, ed. A. Kröner, pp. 1-24, Springer Verlag, Berlin (1984)

J.T. Wasson, Reply to Edward Anders: A discussion of alternative models for explaining the distribution of moderately volatile elements in ordinary chondrites. Earth Planet. Sci. Lett. **36**, 21-28 (1977)

J.T. Wasson, C.- L. Chou, Fractionation of moderately volatile elements in ordinary chondrites. Meteoritics **9**, 69-84 (1974)

J.T. Wasson, G.W. Kallemeyn, Composition of chondrites. Phil. Trans. Roy. Soc. London **325A**, 535-544 (1988)

J.T. Wasson, C.M. Wai, Explanation for the very low Ga and Ge concentrations in some iron meteorite groups. Nature **261**, 114-116 (1976)

D. Welz, P. Deppe, W. Schaefer, H. Sabrowsky, M. Rosenberg, Magnetism of iron-sulfur tetrahedral frameworks in compounds with thallium. I. Chain structures. J. Phys. Chem. Solids **50**, 297-308 (1989)

G. Witt-Eickschen, H. Palme, H.St.C. O’Neill, C.M. Allen, The geochemistry of the volatile trace elements As, Cd, Ga, In and Sn in the Earth’s mantle: New evidence from in situ analyses of mantle xenoliths. Geochim. Cosmochim. Acta **73**, 1755-1778 (2009)

B.J. Wood, D.J. Smythe, T. Harrison, The condensation temperatures of the elements: a reappraisal. Am. Min. **104**, 844–856 (2019)

H. Xu, A. Navrotsky, M.L. Balmer, Y. Su, E.R. Bitten, Energetics of substituted pollucites along the $CsAlSi_2O_6$ – $CsTiSi_2O_{6.5}$ join: A high-temperature calorimetric study. J. Am. Ceram. Soc. **84**, 555-560 (2001)

S. Yoneda, L. Grossman, Condensation of CaO-MgO-$Al_2O_3$-$SiO_2$ liquids from cosmic gases. Geochim. Cosmochim. Acta **59**, 3413-3444 (1995)

L. Yuan, G. Steinle-Neumann, Earth’s missing chlorine may be in the core. J. Geophys. Res. 129 e2023JB027731 (2023)

F.J. Zurfluh, B.A. Hofmann, E. Gnos, U. Eggenberger, “Sweating meteorites” – Water-soluble salts and temperature variation in ordinary chondrites and soil from the hot desert of Oman. Meteorit. Planet. Sci. 48, 1958-1980 (2013)

M. Zabel, K.-J. Range, Ternäre phase im system Eisen-Thallium-Schwefel. Z. Naturforsch. **34b**, 1-6 (1979)